\documentclass[10pt]{article}

\usepackage{amsmath}
\usepackage{amsfonts}
\usepackage{amssymb}
\usepackage{caption}
\usepackage{subcaption}

\usepackage{graphicx}
\usepackage{tikz}\usepackage{pifont}
\usepackage{bbm}
\newcommand{\iu}{\ensuremath{\mathbbm{i}}}
\newcommand{\rn}{\ensuremath{\mathbbm{R}}}

\usepackage{listings}
\usepackage{hyperref}
\usepackage{inconsolata}
\usepackage{tabularx}
\usepackage{color}
\usepackage[left=2.0cm,right=2.0cm,top=1.4cm,bottom=2.0cm]{geometry}

\definecolor{mygray}{rgb}{0.4,0.4,0.4}

\usetikzlibrary{positioning}
\tikzstyle{innerloopstyle} = [draw=black,line width=1pt, minimum width=17cm, minimum height=0.5cm, rounded corners=5pt,align=flush left, text width=16cm,node distance=0.5cm,scale=0.9,line width=0.05cm]
\tikzstyle{outerloopstyle} = [top color=green!50, bottom color=green!10,draw=black,line width=1pt, minimum width=4cm, minimum height=0.5cm, rounded corners=5pt,align=left,node distance=0.5cm,scale=0.9,line width=0.05cm]
\tikzstyle{picstyle} = [align=flush left,node distance=0.1cm]
\tikzstyle{line}=[-latex,draw=black,line width=2pt,line cap=round]
\tikzstyle{line3}=[-latex,draw=gray,line width=2pt,line cap=round]
\tikzstyle{line2}=[-,draw=black,line width=2pt,line cap=round]
\tikzstyle{line4}=[-latex,draw=black,line width=2pt,line cap=round]

\title{ATUS-PRO: A FEM-based solver for the time-dependent and stationary Gross-Pitaevskii equation}
\author{\v{Z}elimir Marojevi\'{c}, Ertan G\"{o}kl\"{u} and Claus L\"{a}mmerzahl\\
ZARM, University Bremen, Am Fallturm, 28359 Bremen, Germany}

\begin{document}

\maketitle

\begin{abstract}
ATUS-PRO is a solver-package written in C++ designed for the calculation of numerical solutions of the stationary- and the time dependent Gross--Pitaevskii equation for local two-particle contact interaction utilising finite element methods. These are implemented by means of the deal.II library \cite{dealII82,BangerthHartmannKanschat2007}. The code can be used in order to perform simulations of Bose-Einstein condensates in gravito-optical surface traps, isotropic and full anisotropic harmonic traps, as well as for arbitrary trap geometries. A special feature of this package is the possibility to calculate non-ground state solutions (topological modes, excited states) \cite{marojevic_energy_2013,yukalov_non-ground-state_1997, yukalov_resonant_2004} for an arbitrarily high non-linearity term. The solver-package is designed to run on parallel distributed machines and can be applied to problems in one, two, or three spatial dimensions with axial symmetry or in Cartesian coordinates. The time dependent Gross--Pitaevskii  equation is solved by means of the fully implicit Crank-Nicolson method, whereas stationary states are obtained with a modified version based on  our own constrained Newton method \cite{marojevic_energy_2013}. The latter method enables to find the excited state solutions.

\end{abstract}

\tableofcontents

\section{Introduction}
In recent years, since the first experimental realisation in 1995, Bose-Einstein condensates (BEC) have caught increased attention in the experimental as well as in the theoretical physics community. The properties of BECs can be well described by means of a partial differential equation of the type ``nonlinear Schr\"odinger equation'' (NLSE). Here this is the Gross-Pitaevskii equation (GPE). In this description, the BEC is treated as a non-uniform, interacting Bose gas at zero temperature. The term ``interacting'' refers in the GPE-description to at least two-particle interactions, which are modelled in a mean-field approximation and give rise to a non-linear term. 

There exist only a few analytic solutions of the GPE, like for example soliton solutions in one dimension, the frequently used Thomas-Fermi approximation for ground states and variational approaches \cite{nicolin_qgaussian_2008}. Also on basis of the Thomas-Fermi approximation the dynamics of BECs in time-dependent traps can be described by means of the scaling approach \cite{castin_scalingapproach}.
However, the approximations and assumptions made in these approaches are not strictly fulfilled in certain situations. Especially when it comes to high-precision measurements, deviations from idealised setups have to be taken into account. For example, the Universality of Free Fall (UFF) is to be tested with quantum matter \cite{schlippert_dualspeciesUFF, muentinga_becinterferometry, goeklue_micgrav} where possible violation signals would be of extremely low magnitude. Since only by including all relevant error sources in the calculations, one is able to distinguish these signals from systematic effects introduced by the experimental equipment or, in general, by the environment.
For this, three-dimensional simulations have to be performed that naturally lead to a high demand of computational power. Often, one needs to utilise massively parallel computer systems.

For example, a comprehensive program package for the simulation of the Gross-Pitaevskii equation has been developed by Muruganandam et al. \cite{Muruganandam20091888} and its extension to C by Vudragovi\'c et al. \cite{Vudragovic20122021,kumar_fortran_2015}. It allows to solve the stationary and non-stationary case for 1D, 2D and 3D for fully anisotropic traps.

In our recent investigation \cite{marojevic_energy_2013} we developed a contrained Newton method in C++ by which numerical solutions for excited states in gravitational surface traps and harmonic traps can be obtained. This was implemented for one-dimensional stationary problems using finite differences. 

Concerning physical motivation to study excited states it has to be mentioned that B\"ucker and coworkers demonstrated experimentally the vibrational state inversion of a Bose-Einstein condensate \cite{buecker_excited-states1, buecker_excited-states2}. The experiments were guided by simulations based on optimal control theory which is a dynamical method. There, the toolbox OCT-BEC  \cite{Hohenester2014194} was used in order to get experimental prescriptions how to control the time-dependent trapping potential. By means of this procedure, the first excited state could be reached. 
As a test of how particular higher order excited states presented in \cite{marojevic_energy_2013} could be realised experimentally, we performed some preliminary calculations utilising the OCT-BEC package. We were able to specify time dependent external potential configurations from which we obtained the excited states up to the third mode. This shows that the predicted states could be produced, in principal, experimentally. The results will be published elsewhere. 

Excited states also find applications in the frame of exploring the interaction of quantum matter with gravity. Energy eigenstates of ultracold neutrons in a gravitational trap have been investigated in order to test Newton`s inverse-square law at small distances by Abele et al. \cite{abele_quantum_2003}. In their setup neutrons fall in the linear gravitational potential and bounce off a neutron mirror.  The neutron flux is measured as a function of the absorber height above the mirror. Therefore, it is a function of the vertical discrete Energy component $E_n$ and thus depends on the mode number $n$. By means of this,  a constraint in the range between $1\, \mu$m and $10 \, \mu$m could be obtained for Yukawa-like potentials. As a side remark, such setups are also known as ``atom trampolines'' or ``quantum bouncers'' whose solutions for the linear case, i.e. the Schr\"odinger equation, are given by Airy functions \cite{wallis_trapping_1992}.  Furthermore, it has been suggested to use quantum bouncers in order to complement tests of the Universality of free fall , see \cite{kajari_uffquantum} and references therein. The main conclusion is that experiments which investigate the dynamics of wave packets, independently of the complexity of the initial state, always probe the ratio $m_g/m_i$ of the gravitational $m_g$ and inertial mass $m_i$. However, the probability density can depend on the product of both masses like $(m_g m_i)^{1/3}$ or other combinations. Moreover, the discrete energy spectrum can behave according to the scaling law $(m^2_g/m_i)^{1/3}$. This could be explored, in principle, by means of spectroscopic methods. This would add to existing experimental procedures and setups for comparing $m_g$ and $m_i$. 

Here we present a highly improved version of our previously mentioned C++ code which has been extended to two- and three-dimensional setups and comes with the following features: (i.) calculation of ground state solutions, as well as (ii.) stationary excited states, (iii.) dynamics simulations via real-time propagation,  and all calculations are based on (iv.) finite element methods (FEM). By the latter, the code can now be used  as a basis for solving problems of higher spatial complexity on massively parallel computer systems (of course also on single multi-core systems), e.g. time propagation of highly excited states in anisotropic time-dependent anharmonic traps.

\section{Physical model and notes on the algorithm }
\subsection{The model}
The physical properties of Bose-Einstein condensates are well characterised by means of the Gross-Pitaevskii equation which in its time-dependent and dimensionless form can be written as 
\begin{equation}
\iu \partial_t \Psi(\vec{x},t) = \left( -\Delta  + V(\vec{x}) + \gamma \vert\Psi(\vec{x},t)\vert^2 \right) \Psi(\vec{x},t). \label{gpe:dyn}
\end{equation}
It describes the dynamics of a many body bosonic system with local self interaction at zero temperature. For this equation to be a good approximation, it is assumed that the s-wave scattering length is much less than the mean interparticle spacing. The wave function $\Psi(\vec{x},t)$ is complex and $\vert\Psi(\vec{x},t)\vert^2$ is interpreted as the local density.  The local self interaction strength is given by $\gamma \in \rn$ (which depends on the s-wave scattering length) and its sign (positive or negative) determines whether the particles of the condensate repel or attract each other. Usually, the external potential $V(\vec{x})$ models a trap in order to spatially confine the condensate, but can also account for e.g. external perturbations on the system. Here we assume that it is bounded from below, i.e. $V(\vec{x})\geq 0$. 

By means of the following ansatz
\begin{equation}
\Psi(\vec{x},t) = \phi(\vec{x}) \exp\left( \iu \mu t \right), \, \phi(\vec{x}) \in \rn, 
\end{equation}
space and time are separated and one obtains the stationary Gross--Pitaevskii equation:

\begin{equation}
\left( -\Delta  + V(\vec{x})-\mu + \gamma \phi^2(\vec{x})\right) \phi(\vec{x})=0, \label{gpe:stat}
\end{equation}
where $\mu$ is the dimensionless chemical potential.  Note that all stationary solutions are real. 

As default setting of the \texttt{atus-pro} package, a gravito-optical surface trap (GOST) is utilised in two dimensions.
That is, the gravitational attraction is superimposed by a harmonic trap, i.e.
\begin{equation}
V(x,y) = \beta x + \omega^2 y^2, \label{GOST}
\end{equation}
for Cartesian coordinates and 
\begin{equation}
V(r,z) = \omega^2 r^2 + \beta z, \label{GOST_cs}
\end{equation}
for the axially symmetrical case, where $r=\sqrt{x^2+y^2}$.
For the three-dimensional case the potential of the GOST is implemented for Cartesian coordinates according to
\begin{equation}
V(x,y,z) =  \beta x + \omega_y^2 y^2 + \omega_z^2 z^2 \label{HO_3D}.
\end{equation}
Inserting an infinite repulsive potential wall at $x=0$ or $z=0$ establishes the boundary condition at the bottom of the trap, whereas $\phi(r=R)=0$. This ensures that the Hamiltonian is bounded from below. Then the potentials (\ref{GOST_cs}) and (\ref{HO_3D}) model the behaviour of a bouncing quantum system,  i.e. the previously mentioned  ``atom trampolines'' or ``quantum bouncers''.   
In our package fully anisotropic harmonic traps are also predefined, that is
\begin{equation}
V(x,y) =  \omega_x^2 x^2+ \omega_y^2 y^2 \label{HO_2D},
\end{equation}
and 
\begin{equation}
V(x,y,z) =  \omega_x^2 x^2+ \omega_y^2 y^2 + \omega_z^2 z^2. 
\end{equation}
The dimensionless parameters $\beta$ and $\omega, \omega_x, \omega_y , \omega_z$ represent the gravitational acceleration and the angular frequencies of the harmonic trap.

\subsection{The contrained Newton method for the stationary case}\label{notes:stat_case}
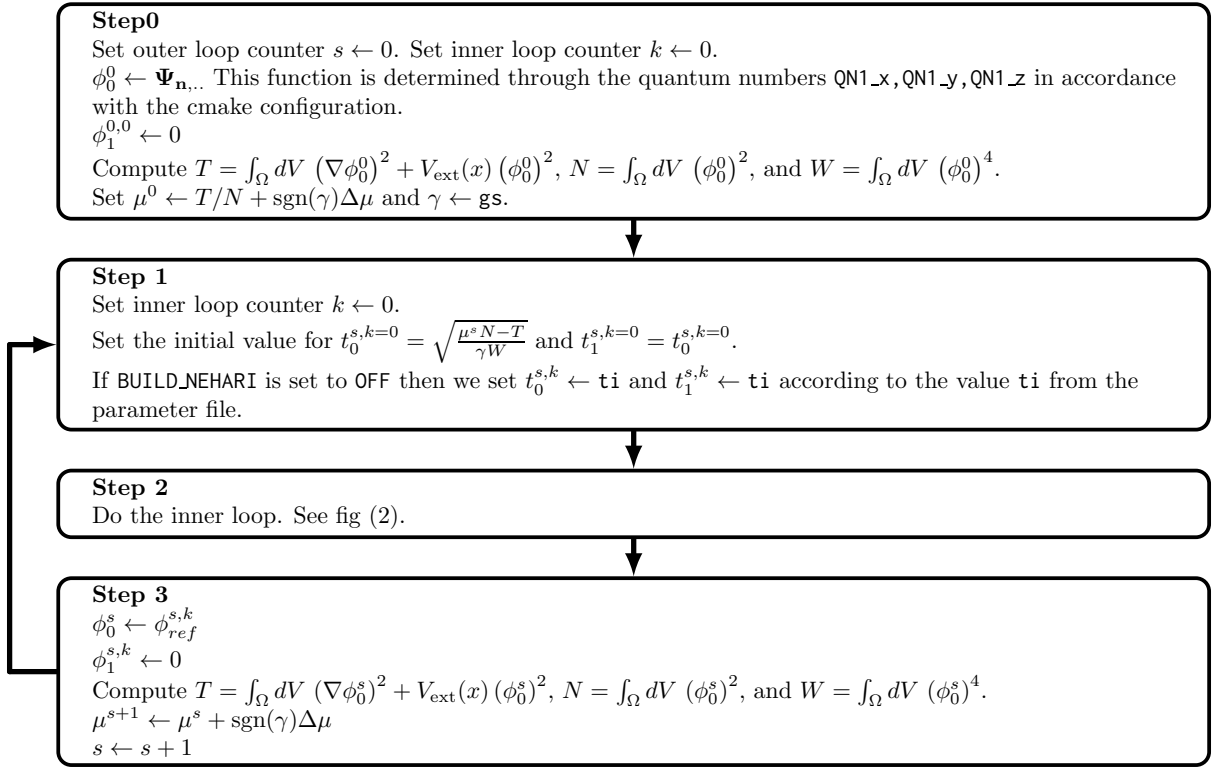
\begin{figure}[h]
\begin{center}
\begin{tikzpicture}[scale=1.0,node distance=0.5cm]

\node[innerloopstyle] (step1) {
{\bf Step0} \\ 

Set outer loop counter $s\leftarrow 0$. Set inner loop counter $k\leftarrow 0$. \\
$\phi_0^0 \leftarrow {\bf \Psi_{n,..} }$ This function is determined through the quantum numbers \texttt{QN1\_x,QN1\_y,QN1\_z} in accordance with the cmake configuration. \\
$\phi_1^{0,0} \leftarrow 0$ \\

Compute 
$ T = \int_{\Omega} dV \, \left( \nabla \phi_0^0 \right)^2 + V_{\rm ext}(x) \left( \phi_0^0 \right)^2  $, $ N = \int_{\Omega} dV \, \left( \phi_0^0\right)^2 $, and $ W = \int_{\Omega} dV \, \left( \phi_0^0 \right)^4 $. 

Set $\mu^0 \leftarrow T/N + \text{sgn}(\gamma) \Delta \mu $ and $\gamma \leftarrow \texttt{gs}$.
};

\node[innerloopstyle] (step2) [below=of step1] {
{\bf Step 1} \\ 
Set inner loop counter $k\leftarrow 0$. \\
Set the initial value for $ t^{s,k=0}_0=\sqrt{\frac{\mu^s N - T}{\gamma W}} $ and $ t_1^{s,k=0}=t_0^{s,k=0} $. \\
If \texttt{BUILD\_NEHARI} is set to \texttt{OFF} then we set $t_0^{s,k} \leftarrow \texttt{ti}$ and $t_1^{s,k} \leftarrow \texttt{ti}$ according to the value \texttt{ti} from the parameter file.
};

\node[innerloopstyle] (step3) [below=of step2] {
{\bf Step 2} \\ 
Do the inner loop. See fig (\ref{sec:alg:flow_chart_inner_loop}).
};

\node[innerloopstyle] (step4) [below=of step3] {
{\bf Step 3} \\ 
$ \phi_0^s \leftarrow \phi^{s,k}_{ref}$ \\
$ \phi_1^{s,k} \leftarrow 0$ \\
Compute 
$ T = \int_{\Omega} dV \, \left( \nabla \phi_0^s \right)^2 + V_{\rm ext}(x) \left( \phi_0^s \right)^2  $, $ N = \int_{\Omega} dV \, \left( \phi_0^s \right)^2 $, and $ W = \int_{\Omega} dV \, \left( \phi_0^s \right)^4 $. 
$ \mu^{s+1} \leftarrow \mu^s + \text{sgn}(\gamma) \Delta \mu$ \\
$ s\leftarrow s+1$
};

\node (help1) [left=0.5cm of step2] {};
\node (help2) [left=0.5cm of step4] {};

\path[line] (step1) -- (step2);
\path[line] (step2) -- (step3);
\path[line] (step3) -- (step4);
\path[line2] (step4) -- (help2.center);
\path[line2] (help2.center) -- (help1.center);
\path[line] (help1.center) -- (step2);

\end{tikzpicture}
\end{center}
\caption{Enhanced Newton method: Flow chart of the outer loop. $s$ is outer loop counter. $k$ is the inner loop counter.} \label{sec:alg:flow_chart_outer_loop}
\end{figure}

\begin{figure}[h]
\begin{center}
\begin{tikzpicture}[scale=1.0,node distance=0.5cm]

\node[innerloopstyle] (step1) {
{\bf Step0} \\ 
Assemble the right hand side \\
\begin{equation*}
F_i^{s,k} = \sum_{K,q} w_q^K \left( \nabla \phi_{ref}^{s,k}(x_q^K) \nabla h_i(x_q^K) + (V-\mu^s) \phi_{ref}^{s,k}(x_q^K) h_i(x_q^K) + \gamma \left( \phi_{ref}^{s,k}(x_q^K) \right)^3 h_i(x_q^K) \right)
\end{equation*}
Compute the initial residual $ \text{res} = \sqrt{\sum_K \Vert F^{s,k} \Vert_{L_2(K)} }.$
};

\node[innerloopstyle] (step2) [below=of step1] {
{\bf Step 1} \\ 
Find the Newton search direction
\begin{equation*}
J_{ij}^{s,k} = \sum_{K,q} w_q^K \left( \nabla h_i(x_q^K) \nabla h_j(x_q^K) + (V-\mu^s) h_i(x_q^K) h_j(x_q^K) + 3\gamma \left(\phi_{ref}^{s,k}(x_q^K) \right)^2 h_i(x_q^K) h_j(x_q^K) \right) 
\end{equation*}
};

\node[innerloopstyle] (step3) [below=of step2] {
{\bf Step 2} \\ 
\begin{equation*}
J^{s,k}_{ij} d^{s,k}_j = F_i^{s,k}
\end{equation*}
};

\node[innerloopstyle] (step4) [below=of step3] {
{\bf Step 3} \\ 
\begin{equation*}
\phi_1^{s,k+1} = \phi_1^{s,k} - \text{sgn}(t_1^{s,k}) d^{s,k}
\end{equation*}
};

\node[innerloopstyle] (step5) [below=of step4] {
{\bf Step 4} \\
Solve the following sytem of equations for $t_0^{s,k+1}$ and $t_1^{s,k+1}$. \\ 
\begin{align*}
0 &= \sum_{K,q} w_q^K \left[ \nabla\phi_{ref}^{s,k+1}(x_q^K) \nabla\phi_0(x_q^K) + (V-\mu^s) \phi_{ref}^{s,k+1}(x_q^K)\phi_0(x_q^K) + \gamma \left(\phi_{ref}^{s,k+1}(x_q^K)\right)^3 \phi_0(x_q^K) \right] \\ 
0 &= \sum_{K,q} w_q^K \left[ \nabla\phi_{ref}^{s,k+1}(x_q^K) \nabla\phi_1^{s,k+1}(x_q^K) + (V-\mu^s) \phi_{ref}^{s,k+1}(x_q^K)\phi_1^{s,k+1}(x_q^K) + \gamma \left(\phi_{ref}^{s,k+1}(x_q^K)\right)^3 \phi_1^{s,k+1}(x_q^K) \right]
\end{align*}
};

\node[innerloopstyle] (step6) [below=of step5] {
{\bf Step 5} \\
$\phi^{s,k+1}_{ref} \leftarrow t_0^{s,k+1} \phi_0 + t_1^{s,k+1} \phi_1^{s,k+1}$ \\
Assemble right hand side $F^{s,k+1}_i$ and compute the residual res. \\
If $\text{res} < \epsilon$ than exit inner loop or else $k\leftarrow k+1$ and go to {\bf Step 1}
};

\node (help1) [left=0.5cm of step2] {};
\node (help2) [left=0.5cm of step6] {};

\path[line] (step1) -- (step2);
\path[line] (step2) -- (step3);
\path[line] (step3) -- (step4);
\path[line] (step4) -- (step5);
\path[line] (step5) -- (step6);
\path[line2] (step6) -- (help2.center);
\path[line2] (help2.center) -- (help1.center);
\path[line] (help1.center) -- (step2);

\end{tikzpicture}
\end{center}
\caption{Enhanced Newton method: Flow chart of the inner loop. $s$ is the outer loop counter. $k$ is the inner loop counter.} \label{sec:alg:flow_chart_inner_loop}
\end{figure}
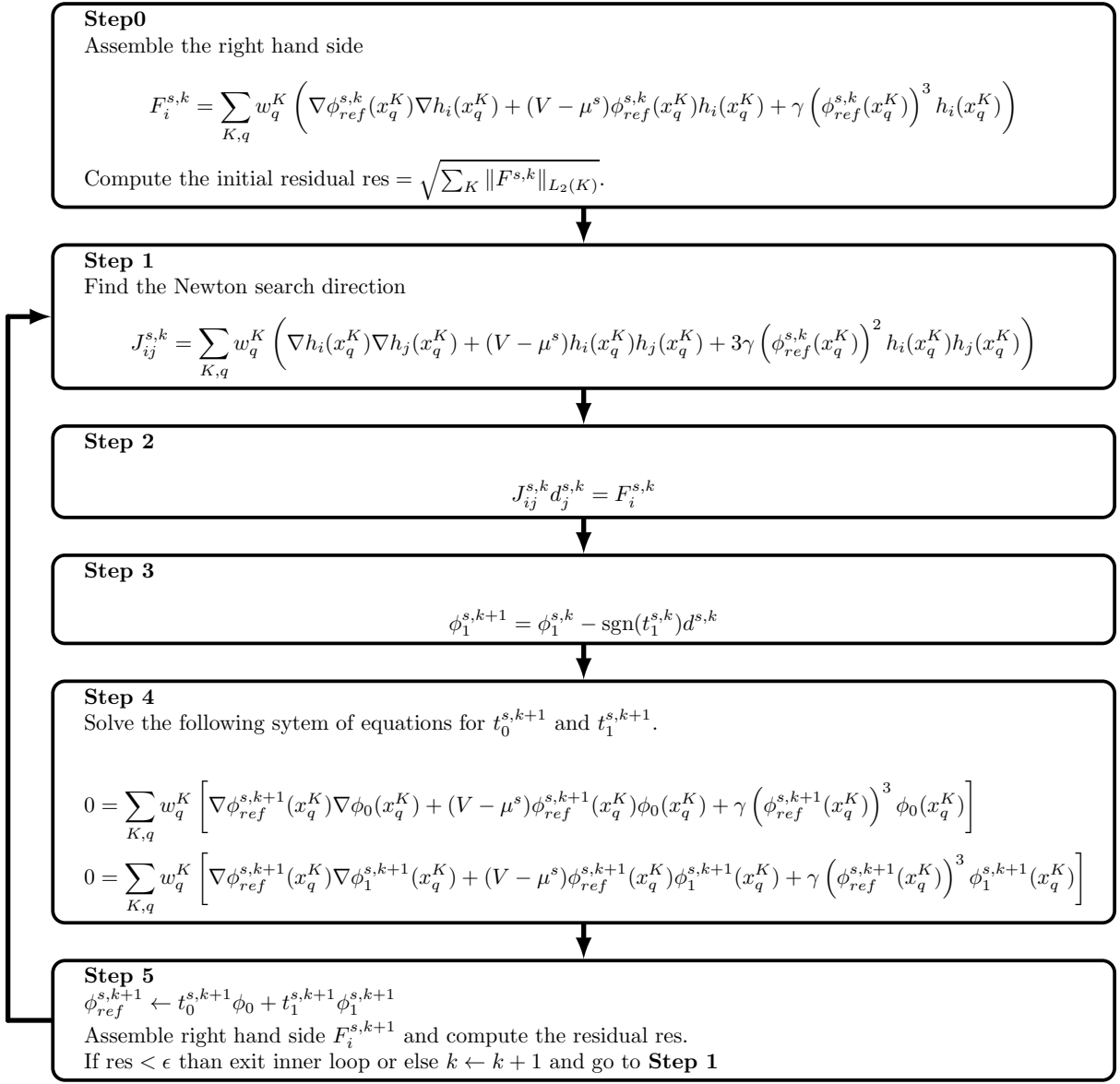

Solutions to the stationary Gross-Pitaevskii equation are obtained by means of a constrained Newton algorithm which is a slightly modified version of the algorithm presented in \cite{marojevic_energy_2013}. In our original implementation which was used to obtain the results in \cite{marojevic_energy_2013}, we used the Finite Difference Method (FDM) for spatial discretisation. However, in order to release a public version of this code we decided to reimplement the algorithm using more sophisticated methods. In particular, we make use of libraries for Finite Element Methods (FEM). The flow charts in figure \ref{sec:alg:flow_chart_outer_loop} and figure \ref{sec:alg:flow_chart_inner_loop} explain how the stationary solutions by means of the \texttt{breed} programs are calculated.

We start reformulating the problem by means of multiplying the stationary GPE (\ref{gpe:stat}) with a test function $h(\vec{x}) \in H^1_0(\Omega)$ and integrating over the volume  $\Omega$, 
\begin{equation}
 \int_{\Omega} \left( -\Delta  + V(\vec{x})-\mu + \gamma \phi^2(\vec{x})\right) \phi(\vec{x}) h(\vec{x}) dV = 0,
\end{equation}
which by means of Green`s identity becomes 
\begin{equation}
 \int_{\Omega} \left( \nabla \phi(\vec{x}) \cdot \nabla h(\vec{x}) + \left(V(\vec{x})-\mu\right)\phi(\vec{x}) h(\vec{x}) + \gamma \phi^3(\vec{x}) h(\vec{x}) \right) dV = 0  \label{gpe:stat-weak}. 
\end{equation}
Note that terms at the boundary $\partial \Omega$ vanish because of the fact that $h(\vec{x})|_{\partial \Omega}$ is zero. The expression (\ref{gpe:stat-weak}) is also known as the \textit{weak formulation of the partial differential equation} (\ref{gpe:stat}). 
Solutions $\phi(\vec{x})$ are critical points of the corresponding functional 
\begin{equation}
A[\phi; \mu]= \int_{\Omega} \left(\left(\nabla \phi(\vec{x})\right)^2+\frac{1}{2} \left(V(\vec{x})-\mu\right)\phi^2(\vec{x})+\frac{\gamma}{4}\phi^4(\vec{x})\right) dV \label{gpe:functional}
\end{equation}
and determined in our algorithm by solving equation (\ref{gpe:stat-weak}) instead of the stationary GPE (\ref{gpe:stat}).
That means we look for the zeroes of (\ref{gpe:stat-weak}) which are not unique. In fact, for $\mu \rightarrow \infty$ one gets infinitely many zeros \cite{rabinowitz_global_1971,rabinowitz_note_1975,fadell_bifurcation_1977,rabinowitz_bifurcation_1977,kielhofer_bifurcation_1988}. However, for a fixed $\mu$ there exist only a finite number of zeroes. In contrast, if we fix the particle number then we have a discrete spectrum of infinitely many different solution. The solution with the smallest $\mu$ is then called ``ground state'' of the system and belongs to a minimum. All other solutions are min-max saddle points, like in the linear case, too. The excited states, or non-ground state solutions, are of that kind.

Before starting the Newton iteration loop, the initial wave function is assembled as a linear combination of two functions,
\begin{equation}
 \phi_{ref}^{s,k} = t_0^{s,k} \phi_0^s + t_1^{s,k} \phi_1^{s,k} \label{newton:reffunction},
\end{equation}
where in the first step the coefficients $t_0^{s,k=0}$ and $t_1^{s,k=0}$ are set automatically or by means of a determined value in the parameter file  \texttt{params.prm}. If set to automatic initialisation, $t_0^{s,k=0} = t_1^{s,k=0} = \sqrt{\frac{\mu^s N - T}{\gamma W}}$ (for a better overview, please see the flow chart in figure \ref{sec:alg:flow_chart_outer_loop}). Here and in the following $s$ and $k$ denote iteration counters.  Note that initially the function $\phi_1^{s=0,k=0}$ is set to zero. In contrast, $\phi_0^0$ is chosen either (i.) automatically w.r.t. the desired nodal structure of the solution, or (ii.) the function can be specified explicitly in the parameter file. In the first case, $\phi_0^{0}$ is given by the solution of the corresponding linear problem, i.e. the solution of the Schrödinger equation with the same potential. 

For example, in the case of a three dimensional harmonic trap the potential is completely characterised by means of the frequencies (\texttt{omega\_x,omega\_y,omega\_z}) given in \texttt{params.prm}. In case (i.) $\phi_0^{0}$ is then determined by specifying the quantum numbers (\texttt{QN1\_x, QN1\_y, QN1\_z}) in the same parameter file, which refer to the stationary solution of the Schr\"odinger equation. By means of this, excited state solutions can also be calculated. 
The precise form of the implemented eigenfunctions will not be reproduced here, in fact this can be looked up in the \texttt{doxygen} documentation which is included in the \texttt{atus-pro} program package. 

Concerning case (ii.) all functions and combinations thereof which are defined in the C standard math library are allowed. However, the manually determined $\phi_0^{0}$ has to respect the boundary conditions.

It is important to mention that after the first Newton iteration step the coefficients $t_0^{s,k}$ and $t_1^{s,k}$ are fixed by the constraints (\ref{newton:contraint-1}) and (\ref{newton:contraint-2}). 

The FEM-form of equation (\ref{gpe:stat-weak}) reads
\begin{equation}
F_i^{s,k} = \sum_{K,q} w_q^K \left( \nabla \phi_{ref}^{s,k}(x_q^K) \nabla h_i(x_q^K) + (V-\mu^s) \phi_{ref}^{s,k}(x_q^K) h_i(x_q^K) + \gamma \left( \phi_{ref}^{s,k}(x_q^K) \right)^3 h_i(x_q^K) \right) \label{newton:deriv_stat_GPE}.
\end{equation}
The integral over the volume translates into a double sum over all cells $K$ and all quadrature points $x^K_q$ in each cell times a weight $w^K_q$, given by the product between the Jacobian of the cell and the weight of the quadrature formula. 
Here the indices are: $i$ denotes the degrees of freedom (in FEM language these are the expansion coefficients in front of the shape functions), $K$ is the sum over all cells and $q$ signifies the sum over all quadrature points in each cell. $h_i(x_q^K)$ is the shape function belonging to the $i$-th degree of freedom.

From this the Jacobian is derived by means of taking the first functional derivative, hence
\begin{equation}
J_{ij}^{s,k} = \sum_{K,q} w_q^K \left( \nabla h_i(x_q^K) \nabla h_j(x_q^K) + (V-\mu^s) h_i(x_q^K) h_j(x_q^K) + 3\gamma \left(\phi_{ref}^{s,k}(x_q^K) \right)^2 h_i(x_q^K) h_j(x_q^K) \right).
\end{equation}
This is needed in order to calculate the search direction $d^{s,k}_j$ of the algorithm for the Newton method, namely
\begin{equation}
 J^{s,k}_{ij} d^{s,k}_j = F_i^{s,k}
\end{equation}
must be solved and next the function 
\begin{equation}
\phi_1^{s,k+1} = \phi_1^{s,k}- \text{sgn}(t_1^{s,k}) d^{s,k},
\end{equation}
is calculated. 

It is important to note that the sign of the coefficient $t_1^{s,k}$ is taken into account which is due to the fact  that the problem has $\mathbb{Z}_2$ symmetry.
Since a solution $\phi$ of the Gross-Pitaevskii equation can be replaced by $-\phi$ being also a solution, this leads to a change of sign for the coefficients as well, i.e. $-t_0^{s,k}$ and $-t_1^{s,k}$. In this case it happens that $F_i^{s,k} \rightarrow - F_i^{s,k}$, however $J^{s,k}_{ij}$ maintains its sign. This would induce a sign change of $d^{s,k}_j$ which is compensated by $\text{sgn}(t_1^{s,k})$.  

In the next step of the Newton loop the constraint of our Newton algorithm is specified by the following system of equations 
\begin{align}
\sum_{K,q} w_q^K \left[ \nabla\phi_{ref}^{s,k+1}(x_q^K) \nabla\phi_0^s(x_q^K) + (V-\mu^s) \phi_{ref}^{s,k+1}(x_q^K)\phi_0^s(x_q^K) + \gamma \left(\phi_{ref}^{s,k+1}(x_q^K)\right)^3 \phi_0^s(x_q^K) \right] &= 0, \label{newton:contraint-1}\\
\sum_{K,q} w_q^K \left[ \nabla\phi_{ref}^{s,k+1}(x_q^K) \nabla\phi_1^{s,k+1}(x_q^K) + (V-\mu^s) \phi_{ref}^{s,k+1}(x_q^K)\phi_1^{s,k+1}(x_q^K) + \gamma \left(\phi_{ref}^{s,k+1}(x_q^K)\right)^3 \phi_1^{s,k+1}(x_q^K) \right] &= 0 \label{newton:contraint-2}.
\end{align}
The mathematical meaning of both equations is to obtain a special point $\phi^{s,k}_{ref}$ in the function space where the $L_2$-gradient given by the l.h.s. of time independent Gross-Pitaevski equation (\ref{gpe:stat}) vanishes in the direction of $\phi_0^s$ and $\phi_1^{s,k}$. That is, one follows all these points until  a critical point is found at which all derivatives vanish. 

Upon finding new values $t^{s,k+1}_0$ and $t^{s,k+1}_1$ from the $k$-th step, a new reference function $\phi^{s,k}_{ref}$ is assembled by means of equation (\ref{newton:reffunction}) and the iteration loop starts again until the $L_2$-norm of the $L_2$ gradient reaches a predetermined value.  That is, the $L_2$ norm of the left hand side of equation (\ref{gpe:stat}) should be close to zero. 

A typical simulation run can now be performed in two ways: (i.) the code will compute a large number of solutions - each for a different chemical potential $\mu^s$ in increasing order or (ii.) one single solution for a fixed $\mu$ is calculated. 

In case of (i.) the code will try to find a solution for an initial chemical potential $\mu^s$ and upon successful completion of the Newton iteration loop it proceeds to determining the next solution for a new $\mu^{s+1}=\mu^s + \Delta \mu$. That is, the Newton loop is evoked each time a new $\mu^s$ is fixed and it runs as long as $s<\texttt{Ndmu}$. Here, the step size $\Delta \mu$ is specified in the parameter file \texttt{params.prm}.  We have to emphasise that this way of calculating solutions is well suited for computations similar to those in \cite{marojevic_energy_2013}.

As a remark, solutions $\phi(\vec{x})$ are currently restricted to real functions. Moreover, the reader should be aware of the fact that the solutions are not normalised to one. Due to the one-to-one correspondence between $\mu$ and $\gamma$ (see for example Fig.8 in \cite{marojevic_energy_2013}) the final particle number $N$ is determined by this pair. In order to get solutions normalised to one, this has to be done by hand and the nonlinearity $\gamma$ must be adjusted accordingly, i.e $\gamma \rightarrow \tilde{\gamma} = N \gamma$. This will be addressed in subsection \ref{stat:normalisation}.  

Here, a general statement is at hand with respect to how degenerated states solutions are handled. Excited state solutions are degenerated by rotational symmetry in case of isotropical harmonic trapping. Usually, this would lead to a failure of the standard Newton algorithm. However, since we use a special constraint, this problem can be circumvented and the Jacobian of our Newton algorithm is not degenerated. This translates to the fact that the structure and the positions of the nodes are fixed by the choice of $\phi_0^0$.  Thus, once chosen, in the case of a two-dimensional harmonic trap, the positions of the lines of nodes in the $x-y$-plane will not change in the course of the calculations (see figure \ref{plot:ho}). In order to obtain rotated solutions w.r.t. the previous one, one has to rotate $\phi_0^0$, too.

However if $\mu$ is accidentally chosen in such a way that it corresponds to a bifurcation point, then the Jacobian becomes degenerated. In this situation the program terminates with a note that the matrix is singular.

For a further and more detailed description of the algorithm, we recommend you to consult the \texttt{doxygen} documentation.

\subsection{Notes on the algorithm for the time-dependent solutions}
In order to advance the complex wave function $ \Psi $ from time $ t $ to $ t+\Delta t $ we have implemented the fully implicit Crank-Nicolson method
for the time dependent Gross--Pitaevskii equation. This scheme is unitary in time, conserves the total number of particles and the total energy.

In order to derive the system of equations, we start with the following ansatz
\begin{equation}
 \iu \frac{\Psi(x,t+\Delta t )-\Psi(x,t)}{\Delta t} - \left( -\Delta + V + \gamma \left\vert \frac{\Psi(x,t+\Delta t )+\Psi(x,t)}{2} \right\vert^2 \right) \left(\frac{\Psi(x,t+\Delta t )+\Psi(x,t)}{2}\right) = 0.
\end{equation}

Due to the fact that $\Psi$ is complex, it is split into a real- and imaginary part:

\begin{equation}
 u_t := \text{Re}\,\Psi(x,t+\Delta t),\, v_t := \text{Im}\,\Psi(x,t+\Delta t),\, u := \text{Re}\,\Psi(x,t),\,\text{and } v := \text{Im}\,\Psi(x,t).
\end{equation}

The space and time discretised version of the Gross--Pitaevskii equation then becomes a system of two coupled non linear differential equations for the unknown functions $ u_t $ and $ v_t $.
\begin{equation}
 \vec{F} \left[ u_t, v_t \right] := 
  \begin{pmatrix}
  u_t - u - \frac{\Delta t}{2} \left(-\Delta+V\right) \left(v_t+v\right) - \frac{\Delta t \gamma}{8} \left( \left(v_t+v\right)^2 + \left(u_t+u\right)^2 \right) \left(v_t+v\right) \\
  v_t - v + \frac{\Delta t}{2} \left(-\Delta+V\right) \left(u_t+u\right) + \frac{\Delta t \gamma}{8} \left( \left(v_t+v\right)^2 + \left(u_t+u\right)^2 \right) \left(u_t+u\right) 
  \end{pmatrix} = 0 \text{.} \label{sec:ck:F}
\end{equation}

In order to find the new wave function at time step $t+\Delta t$ we use the standard Newton method (not to be confused with our constrained Newton method). Multiplying both components of (\ref{sec:ck:F}) with two different test functions and integrating over the volume leads to the weak formulation. Carrying out the the FEM discretisation we find for the right hand side

\begin{equation}
F^k_i = \sum_{K,q} w^K_q \begin{pmatrix}
a^- g_i - \frac{\Delta t}{2} \nabla b^+ \nabla g_i + \frac{\Delta t}{2} V b^+ g_i - \frac{\Delta t \gamma}{8} \left( \left(a^+ \right)^2 + \left(b^+\right)^2 \right) b^+ g_i \\
b^- h_i + \frac{\Delta t}{2} \nabla a^+ \nabla h_i + \frac{\Delta t}{2} V a^+ h_i + \frac{\Delta t \gamma}{8} \left( \left(a^+ \right)^2 + \left(b^+\right)^2 \right) a^+ h_i
\end{pmatrix} \label{sec:ck:F_FEM},
\end{equation}
where
\begin{equation}
a^+ := u^k_t(x^K_q) + u^k(x^K_q)\text{, } a^- := u^k_t(x^K_q) - u^k(x^K_q) \text{, } b^+ := v^k_t(x^K_q) + v^k(x^K_q) \text{, } b^- := v^k_t(x^K_q) - v(x^K_q) \text{.}
\end{equation}

\begin{figure}[h]
\begin{center}
\begin{tikzpicture}[scale=1.0,node distance=0.5cm]

\node[innerloopstyle] (step1) {
{\bf Step0} \\ 
Set the loop counter $k \leftarrow 0$. Set the intial guess for $u^0_t \leftarrow 0$ and $v^0_t \leftarrow 0$. \\
Assemble the right hand side 
\begin{equation*}
F^k_i = \sum_{K,q} w^K_q \begin{pmatrix}
a^- g_i - \frac{\Delta t}{2} \nabla b^+ \nabla g_i + \frac{\Delta t}{2} V b^+ g_i - \frac{\Delta t \gamma}{8} \left( \left(a^+ \right)^2 + \left(b^+\right)^2 \right) b^+ g_i \\
b^- h_i + \frac{\Delta t}{2} \nabla a^+ \nabla h_i + \frac{\Delta t}{2} V a^+ h_i + \frac{\Delta t \gamma}{8} \left( \left(a^+ \right)^2 + \left(b^+\right)^2 \right) a^+ h_i
\end{pmatrix}
\end{equation*}
Compute the initial residual $ \text{res} = \sqrt{\sum_K \Vert F^k \Vert_{L_2(K)} }.$
};

\node[innerloopstyle] (step2) [below=of step1] {
{\bf Step 1} \\ 
Assemble the Jacobian
\begin{multline}
J^k_{ij} = \sum_{K,q} w_q^K
\begin{pmatrix}
\nabla g_j(x_q) \\
\nabla h_j(x_q)
\end{pmatrix}^T 
\begin{pmatrix}
0 & -\frac{\Delta t}{2} \\
\frac{\Delta t}{2} & 0
\end{pmatrix} 
\begin{pmatrix}
\nabla g_i(x_q)\\
\nabla h_i(x_q)
\end{pmatrix}
+ \\ w_q^K 
\begin{pmatrix}
g_j(x_q) \\ 
h_j(x_q)
\end{pmatrix}^T 
\begin{pmatrix}
1-\frac{\Delta t \gamma}{4} a^+ b^+ & -\frac{\Delta t}{2} V - \frac{\Delta t \gamma}{4} \left( 3\left(b^+\right)^2 + \left(a^+\right)^2 \right) \\
\frac{\Delta t}{2} V + \frac{\Delta t \gamma}{4} \left( \left(b^+\right)^2 + 3\left(a^+\right)^2 \right) & 1+\frac{\Delta t \gamma}{4} a^+ b^+
\end{pmatrix}  
\begin{pmatrix}
g_i(x_q) \\ 
h_i(x_q)
\end{pmatrix}
\end{multline}
};

\node[innerloopstyle] (step3) [below=of step2] {
{\bf Step 2} \\ 
Compute the Newton search direction via solving the linear system 
$ J^k_{ij} d_j = F^k_i $.
};

\node[innerloopstyle] (step4) [below=of step3] {
{\bf Step 3} \\ 
Do the Newton step $ F^{k+1}_i = F^k_i - d_i  $. \\
$ k \leftarrow k+1 $.
};

\node[innerloopstyle] (step5) [below=of step4] {
{\bf Step 4} \\ 
Assemble the right hand side $ F^k_i $ and compute the residual res of $ F^k_i $.
};

\node[innerloopstyle] (step6) [below=of step5] {
{\bf Step 5} 
If the residual res is smaller than $10^{-16}$, the set $u \leftarrow u^k_t$, $v \leftarrow v^k_t$ and terminate the loop or
else go to {\bf Step 1}.
};

\node (help1) [left=0.5cm of step2] {};
\node (help2) [left=0.5cm of step6] {};

\path[line] (step1) -- (step2);
\path[line] (step2) -- (step3);
\path[line] (step3) -- (step4);
\path[line] (step4) -- (step5);
\path[line] (step5) -- (step6);
\path[line2] (step6) -- (help2.center);
\path[line2] (help2.center) -- (help1.center);
\path[line] (help1.center) -- (step2);

\end{tikzpicture}
\end{center}
\caption{Flow chart of the real time propagation. Here we defined $a^+ := u^k_t(x^K_q) + u^k(x^K_q)$, $a^- := u^k_t(x^K_q) - u^k(x^K_q)$, $b^+ := v^k_t(x^K_q) + v^k(x^K_q)$, $b^- := v^k_t(x^K_q) - v(x^K_q)$.} \label{sec:alg:flow_chart_ck}
\end{figure}
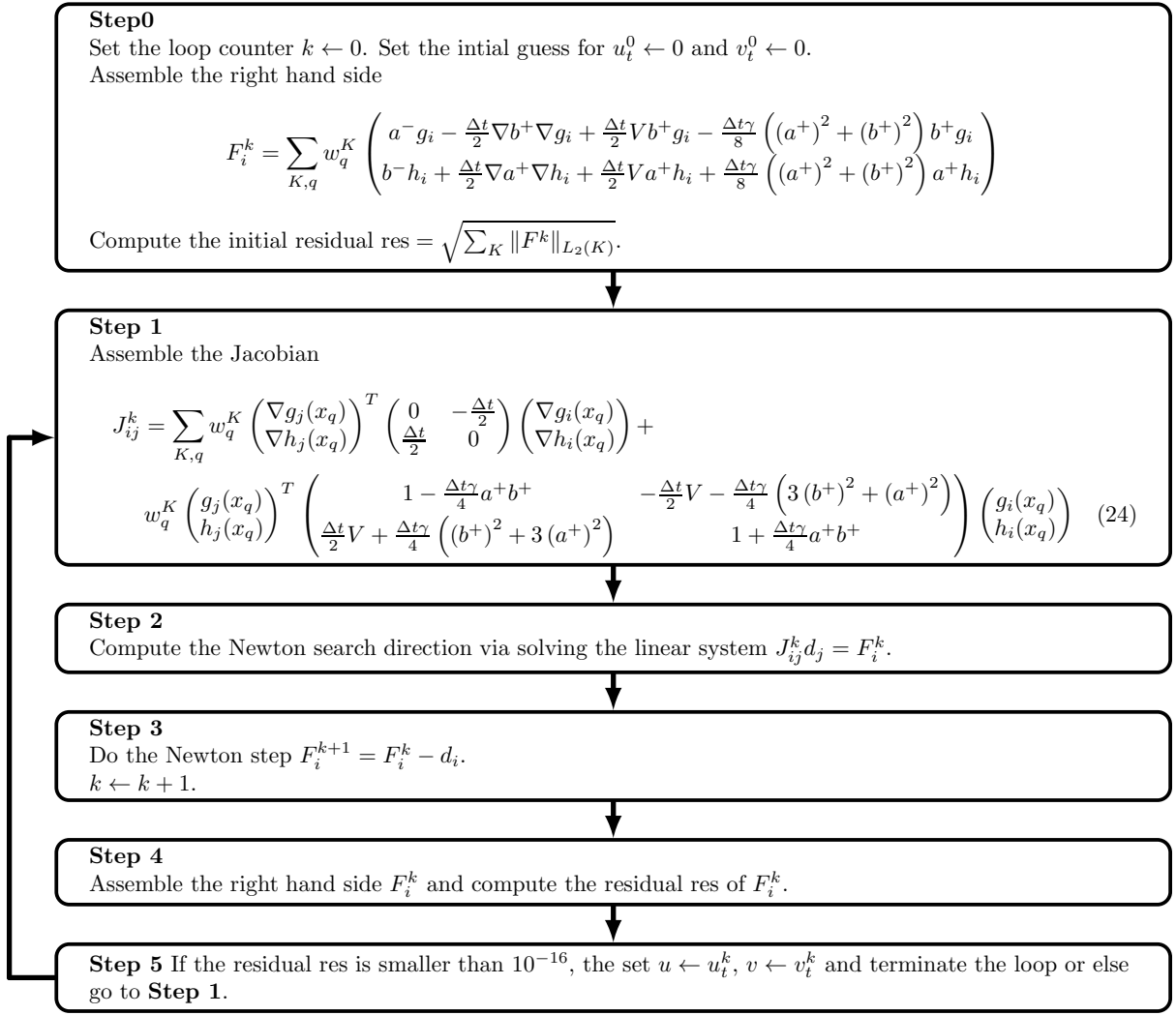

Here, again, we sum over all cells $ K $ and over all quadrature points $ x_q^K $ of each cell. $ w_q^K $ is the weight of the quadrature formula evaluated  at $ x_q^K $ times 
the Jacobian of the unit cell. $ g_i $ are the test functions for the real part and $ h_i $ for the imaginary part evaluated at position $x_q^K$. The reader should be aware that the practical implementation is slightly different than (\ref{sec:ck:F_FEM}) suggests. As a matter of fact, the form here has been chosen for better understanding. In practice, both discretised functions representing the imaginary and real part of the wave function are represented by one vector containing all degrees of freedom.  More information about vector valued problems in the framework of deal.II can be found in the deal.II documentation. The sketch of the whole algorithm can be found in figure \ref{sec:alg:flow_chart_ck}.

\section{Running the program}

\subsection{Available programs}
The whole package consists of eight programs. Six are used for the simulations, three of which are responsible for computations of stationary states, and the remaining three are used for the real-time propagation. The last two programs serve for generating parameter files.

\begin{tabularx}{18cm}{|X|X|}
\hline 
\texttt{breed\_1} & One-dimensional solver for the stationary Gross--Pitaevskii equation.\\
\hline 
\texttt{breed} & Solver for the stationary Gross--Pitaevskii equation in Cartesian coordinates.\\
\hline 
\texttt{breed\_cs} & Same as above but for cylinder symmetry. \\
\hline 
\texttt{rtprop\_1} & One-dimensional solver for the real-time propagation.\\
\hline 
\texttt{rtprop} & Solver for the real-time propagation of the wave function in Cartesian coordinates.\\
\hline 
\texttt{rtprop\_cs} &  Same as above but for cylinder symmetry. \\
\hline 
\texttt{gen\_params} & {Generates a directory with sub-directories containing parameter files } \\
  & for different quantum numbers.\\
\hline 
\texttt{gen\_params\_cs} & {The same like above, just for cylindrical symmetry.} \\
\hline 
\end{tabularx}
\\
For taking advantage of parallelisation, the 2D- and 3D simulations (\texttt{breed}, \texttt{breed\_cs}, \texttt{rtprop} and \texttt{rtprop\_cs}) should be called via \texttt{mpirun -np [no of CPU cores] [program name]}. Of course, single core execution is also possible. Note that in case of the 1D simulations only single core operation is supported.  
When the \texttt{atus-pro} package is compiled for the first time without having modified cmake options beforehand, the programs \texttt{breed}, \texttt{breed\_cs}, \texttt{rtprop} and \texttt{rtprop\_cs} operate in 2D-mode. 
Full 3D mode is supported by \texttt{breed} and \texttt{rtprop}. 

\subsection{Setup of trapping potential and parameters}
The potential $V(x)$ is specified inside the source code where pre-defined traps are: (i.) gravito-optical surface trap and (ii.) harmonic trap.  The desired type of trap and the dimensionality of the problem has to be specified prior to compiling the package via CMake options (see subsection \ref{CMake_options}). This can be changed at any time and requires recompiling the code after having made the changes via the according CMake options. In practice, recompilation just takes a small amount of time. In order to use arbitrary traps, you can define them in the source code of the respective solver. For more details, it is suggested to take a closer look at the \texttt{doxygen} HTML documentation. Note that adjustments of all parameters related to physical or numerical quantities, e.g. the trap frequencies, do not require recompiling the code since this is done through the parameter file.   

\subsection{Parameter file}
All programs expect a parameter file named  \texttt{"params.prm"} where the values of the necessary parameters are set. It must be located in the folder from which the simulation runs are invoked and can be created via the programs themselves upon first invocation from within any directory. The parameter file is then automatically generated with default values and can be edited afterwards. Alternatively, you can use the programs \texttt{gen\_params} or \texttt{gen\_params\_cs} which will generate a folder with sub folders labelled with respect to the quantum numbers of the initial guess functions. Each of those sub folders contains its own parameter file.  This set of cases is meant to be used with \texttt{breed} and \texttt{breed\_cs}. In case that the predefined set of values of the parameters should be changed, editing \texttt{gen\_params.cpp} and \texttt{gen\_params\_cs.cpp} is required, followed by a recompilation of the \texttt{atus-pro} package.

\subsection{On the behaviour of the stationary states solvers: single vs. multiple solutions}\label{single_multiple}
As already mentioned, the default strategy of the solvers \texttt{breed\_1}, \texttt{breed} and \texttt{breed\_cs} is to generate consecutive solutions w.r.t. increasing chemical potential $\mu=\mu_0 + \Delta \mu$. This is the standard setting and cannot be changed via the parameter file. If one is interested in calculating only one stationary solution, then the source files \texttt{breed\_1.cpp}, \texttt{breed.cpp} or \texttt{breed\_cs.cpp} need to be edited. Note that only the source file of the corresponding program, which you like to use for the calculation, has to be modified. In each of the mentioned files, in the \texttt{main} section of the code, the default statement \texttt{solver.run2b()} must be turned into a comment and \texttt{solver.run()} has to be uncommented. This is then followed by a recompilation of the \texttt{atus-pro} package.  

\subsection{Output}
\paragraph{Stationary states}\label{output:stat_states}
When the programs \texttt{breed\_1}, \texttt{breed} and \texttt{breed\_cs} are invoked, the output on the terminal gives you various information about the status of the simulation. In the case of the test-run, which will be discussed in section \ref{TESTRUN}, the first step of the calculation generates an output like this : 
\begin{lstlisting}
- 0 - 0000/  # - 0 - : number of iteration; - 0000/ : subdirectory where the calculation takes place
Solving...
Counter        == 0 # number of iteration 
res            == 4.69515e-07 # L2 norm of L2 gradient (ideally it would be zero). 
resp           == 0.000293776 # difference of res: res(iteration before)-res(current)
|res / resp|   == 0.00159821 # can be used to estimate convergence rate 
mu             == 5.7 # Chemical potential 
gs             == 1 # Self interaction strength 
t1             == 1.34114 # 1st factor for constrained Newton algorithm 
t2             == 0.997539 # 2nd factor for contrained Newton algorithm
l2 norm t      == 1.67145 # L2-norm of vector t=(t1,t2)
total no of cells == 154624 # total number of cells of all nodes
total no of active cells == 115969 # total number of cells being used during integration
\end{lstlisting}
Note that we added short explanations indicated by the comment symbol ``\texttt{\#}''. 
Concerning the output variable \texttt{res}, the L2-gradient is the left hand side of the stationary GPE in equation (\ref{gpe:stat}), i.e. a good solution is characterised by a value close to zero. Here, the default value of the L2-gradient is set to $10^{-5}$. It can be adjusted in the \texttt{params.prm} file by means of the variable \texttt{set epsilon}. 
Upon reaching this value, the calculation for the actual \texttt{mu} is finished and the norm $N$ of the wave function is printed to the terminal. Note that the self interaction strength \texttt{gs} corresponds to the value of $\gamma$ in eq. (\ref{gpe:stat}).  

\paragraph{Dynamics}
Upon starting \texttt{rtprop\_1}, \texttt{rtprop} and \texttt{rtprop\_cs} the terminal outputs typically show the following information:   
\begin{lstlisting}
min_cell_diameter = 0.0828641 
max_cell_diameter = 10.6066
dt/dx^2 == 1.45636 # Stability criterion
t == 0 # Time-stamp of initial state
N == 1.8432 # Number of particles 
p == 0, 0, 0 # Expectation value of momentum (p_x,p_y,p_z) 
pos == 2.74494, -3.61401e-16, 0 # Expectation value of position (x,y,z)
t == 0 # Initial time of simulation 
m_res = 0.000265367 # L2-norm of L2-gradient of time-dependent GPE
m_res = 4.78803e-09
\end{lstlisting}
This case refers to the initial state in the directory \texttt{0000/} for the test-run in Cartesian coordinates.
Note that the user can get a rough idea of the grid in use by means of the values \texttt{min\_cell\_diameter} and \texttt{max\_cell\_diameter}. 
Concerning the stability criterion, which is based on the Courant condition, we would like to emphasise that it is generally recommended to have $dt/dx^2 \lesssim 1/2$. This value can be influenced by adjusting the time step \texttt{dt} in the parameter file \texttt{params.prm}. The default value is \texttt{set dt        = 1e-2} and setting it to \texttt{1e-3} will lead in the presented case to $dt/dx^2 = 0.145636$.

Note that the calculations for each time step are carried out until the residual \texttt{m\_res} reaches $10^{-16}$. This setting can only be adjusted in the program code. Using a smaller value for the residual than the predetermined one is not suggested since it would be below machine epsilon for double precision floating point numbers and leads the programs to perform calculations ad infinitum.    

\paragraph{Files}
All programs store the data in the deal.II native format and in the vtu format. The first format is used for storing the triangulation data and the wave function. For example, you can use a stationary solution from \texttt{breed} as input to the real- time propagation program \texttt{rtprop}. The second file format is meant to be used for graphical post-processing of the simulation data. We tested this extensively under ParaView (\url{http://www.paraview.org/}) and VisIt (\url{https://wci.llnl.gov/simulation/computer-codes/visit/}). Moreover, both software packages are 
able to perform various data post processing tasks.
\\

\textbf{Stationary states:} In the case of stationary states the output is written into the files:
(a) \texttt{final.bin}, \texttt{final.bin.info} and, e.g. \texttt{final-00001.vtu} (two- and three-dimensions), whereas (b) in the one-dimensional case 
these are \texttt{final.bin} and, e.g.  \texttt{final-00001.gnuplot}. Note that in the latter case the graphical output format is best suited for Gnuplot (\url{http://www.gnuplot.info/}).  
These file names are derived from the \texttt{"filename"} parameter entry in  \texttt{"params.prm"}.

Moreover, a result file \texttt{results.csv} is written to the folder from which the simulation has been initiated. 
It contains the following data (only one entry is shown):

\begin{lstlisting}[caption=results.csv]
# mu;gs;N;Counter;status
5.7;1;1.8432;1;0 
\end{lstlisting}

Here, \texttt{mu} is the dimensionless chemical potential, \texttt{gs} is the nonlinearity $\gamma$, \texttt{N} is the norm of the resulting wave function, \texttt{Counter} refers to the number (\texttt{Counter}+1) of iterations which was needed for reaching convergence and finally \texttt{status} indicates whether the calculations have been successful (\texttt{status} = 0) or not (\texttt{status} = 1).

Moreover, the status file \texttt{log.csv} will be created. In the case of single solution computations (see subsection \ref{single_multiple}) it is located in the folder from where the simulation has been started, whereas for multiple solutions, this file is found in each respective sub-directory.  Typically, it has the following entries:

\begin{lstlisting}[caption=log.csv]
# Counter;res;resp;|res / resp|;mu;gs;t1;t2;l2 norm t;total no of cells;total no of active cells
0;4.69514e-07;0.000293776;0.00159821;5.7;1;1.34114;0.997539;1.67145;154624;115969
\end{lstlisting}
These are the same values like those discussed in subsection \ref{output:stat_states}. 
\\

\textbf{Real time propagation:} For the dynamical case the following output files are generated (example: time step $t=0.1$):
(a) \texttt{solution-0.100000.vtu}, while (b) \texttt{solution-0.100000.gnuplot}, and so on.  

Note that to initiate a run of the programs \texttt{rtprop} and \texttt{rtprop\_cs} for two- and three-dimensional problems the files \texttt{final.bin} and \texttt{final.bin.info} are required. In contrast, for the time propagation in one dimension \texttt{rtprop\_1}, no file \texttt{final.bin.info} is needed. In fact, it is not generated by \texttt{breed\_1}. Note that for all real-time simulation programs the parameter file \texttt{params.prm} is needed. 
-
In addition, \texttt{rtprop} and \texttt{rtprop\_cs} create a file named \texttt{solution.pvd} which contains a temporal sequence of the results for intermediate time steps. This applies in case you like to produce a film of the real-time propagation with ParaView.

\begin{figure}[h]
\centering
\includegraphics[width=8cm, keepaspectratio=true]{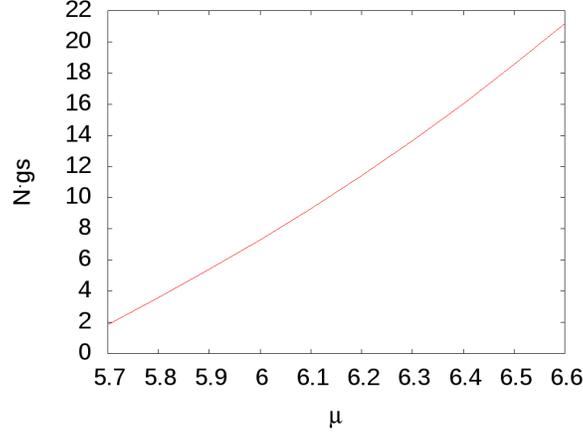}
\caption{\small Rescaled nonlinearity $\tilde{\gamma}=N \cdot \texttt{gs}$ as function of the chemical potential $\mu$. The solutions have been obtained in the testrun (section \ref{TESTRUN}) by the program \texttt{breed} for multiple solutions of the first excited state of a BEC in a GOST. Each $\mu_i$ is eigenvalue to the corresponding solution $\tilde{\phi}_i(\vec{x})$ of the stationary Gross-Pitaevskii equation.\label{plot:mu-gamma}}
\end{figure}

\subsection{Normalisation of the wave function and fixing the particle number}\label{stat:normalisation}
As mentioned earlier in \ref{notes:stat_case}, the resulting wave functions from the stationary solvers are not normalised. In order to normalise them to one, the resulting norm of the wave function $N$ can be used from the file \texttt{results.csv} together with the value for \texttt{gs}. That is, one selects for a particular chemical potential $\mu_i$ the corresponding norm $N_i$. Here, the index $i$ runs from $1$ to the number of the last row in \texttt{results.csv}. Then the resulting nonlinearity can be set to $\gamma_i \rightarrow \tilde{\gamma}_i = N_i \cdot \texttt{gs}$ and the computed wave function for the pair $(\mu_i, N_i)$ must be rescaled according to $\phi_i \rightarrow \tilde{\phi}_i= \phi_i/\sqrt{N_i}$. In terms of the stationary GPE (\ref{gpe:stat}) the new quantities now fulfill the eigenvalue equation
\begin{equation}
\mu_i \tilde{\phi}_i= \left( -\Delta  + V(\vec{x})+ \tilde{\gamma}_i (\tilde{\phi}_i(\vec{x}))^2\right) \tilde{\phi}_i(\vec{x}), \label{gpe:stat_norm}
\end{equation} 
with normalisation 
\begin{equation}
 \int_V{(\tilde{\phi}_i(\vec{x}))^2 d^3x} =1.
\end{equation}
Now suppose that the initial (dimensionless) GPE (\ref{gpe:stat}) is the result of rescaling the (dimensionful) equation with the help of a typical length-scale $L$ of the system, 
\begin{equation}
 \vec{x} \rightarrow L\cdot \vec{x}, \qquad \phi(\vec{x}) \rightarrow \phi(\vec{x}) \cdot L^{-3/2},
\end{equation}
and the rescaled wave function is supposed to be normalised to one. 
In this case the dimensionless non-linearity term reads
\begin{equation}
\gamma_{p}:= \frac{2 m g_s}{\hbar^2 L},  \qquad \text{where } g_s= \frac{4 \pi \hbar^2 a_s}{m},  
\end{equation}
thus 
\begin{equation}
 \gamma_{p} = \frac{8 \pi a_s}{L}.
\end{equation}
Here $a_s$ is the s-wave scattering length and $m$ is the mass of the atom species.
Because of the fact that the computations are made with non-normalised wave functions, $\gamma_{p}$ cannot be identified with the initial nonlinearity $\gamma$ of equation (\ref{gpe:stat}). However, by means of eq. (\ref{gpe:stat_norm}) one can identify the following products:
\begin{equation}
\tilde{\gamma}= \gamma \cdot N = \gamma_p \cdot N_p,
\end{equation}
since the wave function $\tilde{\phi}(\vec{x})$ is normalised to one. Note that $N_p$ is the real, physical value of the particle number. 
By means of this one gets the particle number
\begin{equation}
 N_p = \frac{\tilde{\gamma}}{8 \pi a_s}L \label{gpe:N_phys}.
\end{equation}
Alternatively, there exists the possibility to read the particle number $N_p$ directly from the table in \texttt{results.csv}. One demands 
\begin{equation}
 \gamma = \gamma_p,
\end{equation}
which leads to $N=N_p$ and $\gamma=8 \pi a_s/L$. Upon fixing the length-scale $L$, the value of $\gamma$, that is \texttt{gs}, is determined and can be adjusted in the parameter file \texttt{params.prm}. From now on, the value $N$ in \texttt{results.csv} has the meaning of a real particle number.   

If one wishes to perform simulations for a determined number of particles $N^0_p$, it is suggested to compute first multiple solutions with different $\mu_i$ (see the testrun in section \ref{TESTRUN}). The resulting ($\tilde{\gamma_i}, \mu_i$) pairs can be visualised by plotting the first column versus the product of the second and third column of the file \texttt{results.csv}. In this way one can get an idea about the functional relation between $\tilde{\gamma_i}$ and $\mu_i$ (see figure \ref{plot:mu-gamma}).  By means of equation (\ref{gpe:N_phys}) $\tilde{\gamma}^0$ is determined by the desired particle number $N^0_p$. Finally, the pair ($\mu^0, \tilde{\gamma}^0$) can be read (or interpolated) from the plot.

\subsection{Practical guidelines and hints}
Concerning the stationary solvers, if the external potential $V(x)$ is too shallow, it could happen that the solver ``jumps'' between different solution branches. That means, the outcome of the solution becomes partly unpredictable, i.e. not controllable. Figuratively speaking, the wave function leaves the potential partially and gets delocalised, covering the whole simulation box. Thus the corresponding eigenfunctions are also possible solutions. In order to avoid this, one should scale the trap accordingly in order to get a deeper potential.  

Should the convergence behaviour be problematic, in many cases, it can be handled by modifying two parameters in the file \texttt{params.prm}. We suggest to decrease the increment $\Delta \mu$ via the value of \texttt{dmu} or/and adjust the damping factor \texttt{df} of the contrained Newton algorithm, which is set per default to 1 and should be in the range $\texttt{df} \in [0,1]$.  

As a general statement, try to avoid choosing an unsuitable spatial scale w.r.t. the typical grid scale. This could happen when the GPE is rescaled with a length scale much smaller than the average cell length. It is then highly probable that the solvers won`t converge. In our simulations we usually take $10^{-6}$ m as scaling length. Of course, this depends on the external potential in use.   

\begin{figure}[h]
\begin{subfigure}{5cm} 
\includegraphics[width=5cm, keepaspectratio=true]{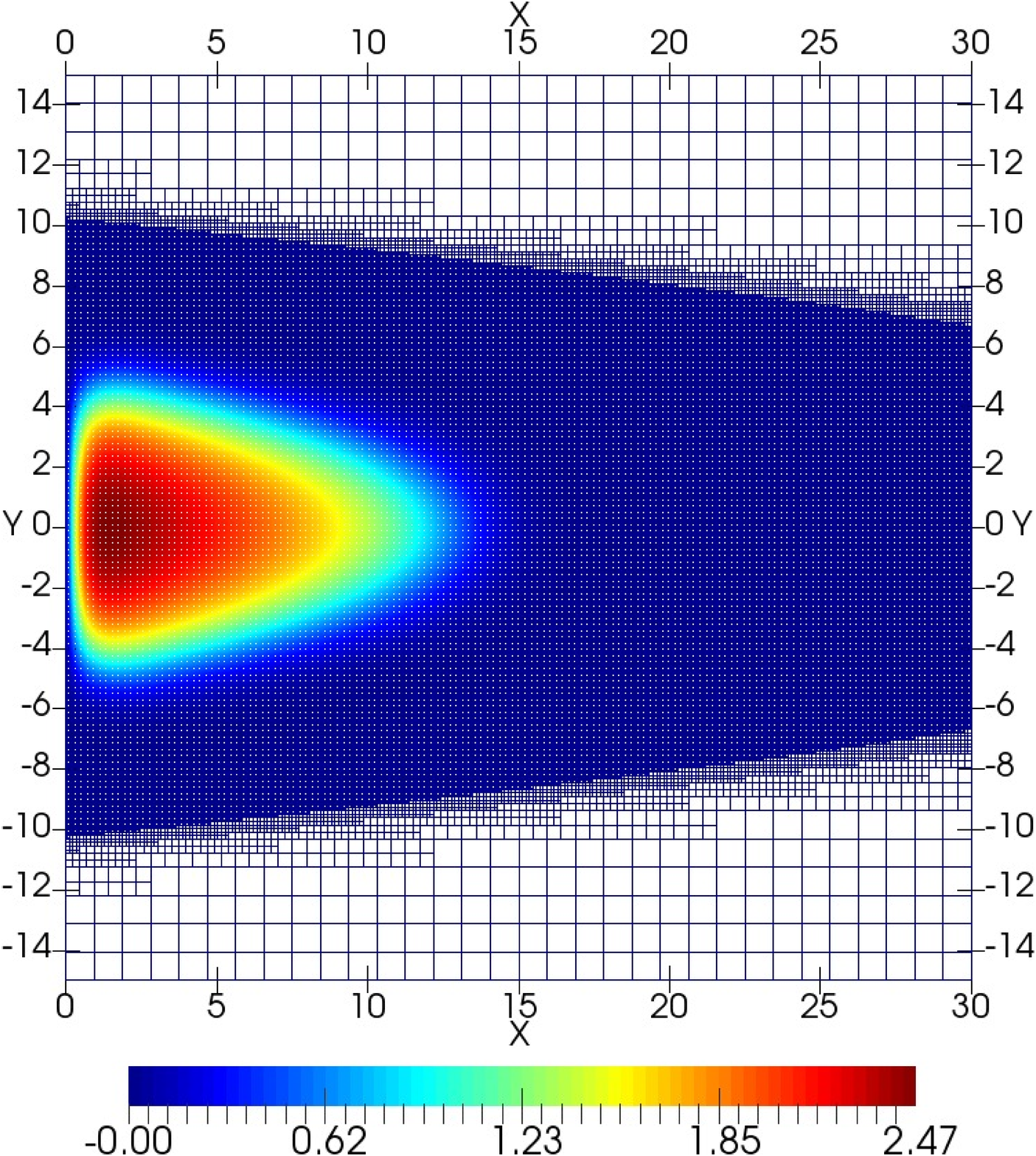}
\caption{$\mu=7$, $N\approx 239.03$, \\ \texttt{QN1\_x=0, QN1\_y=0} }
\end{subfigure}
\qquad
\begin{subfigure}{5cm} 
\includegraphics[width=5cm, keepaspectratio=true]{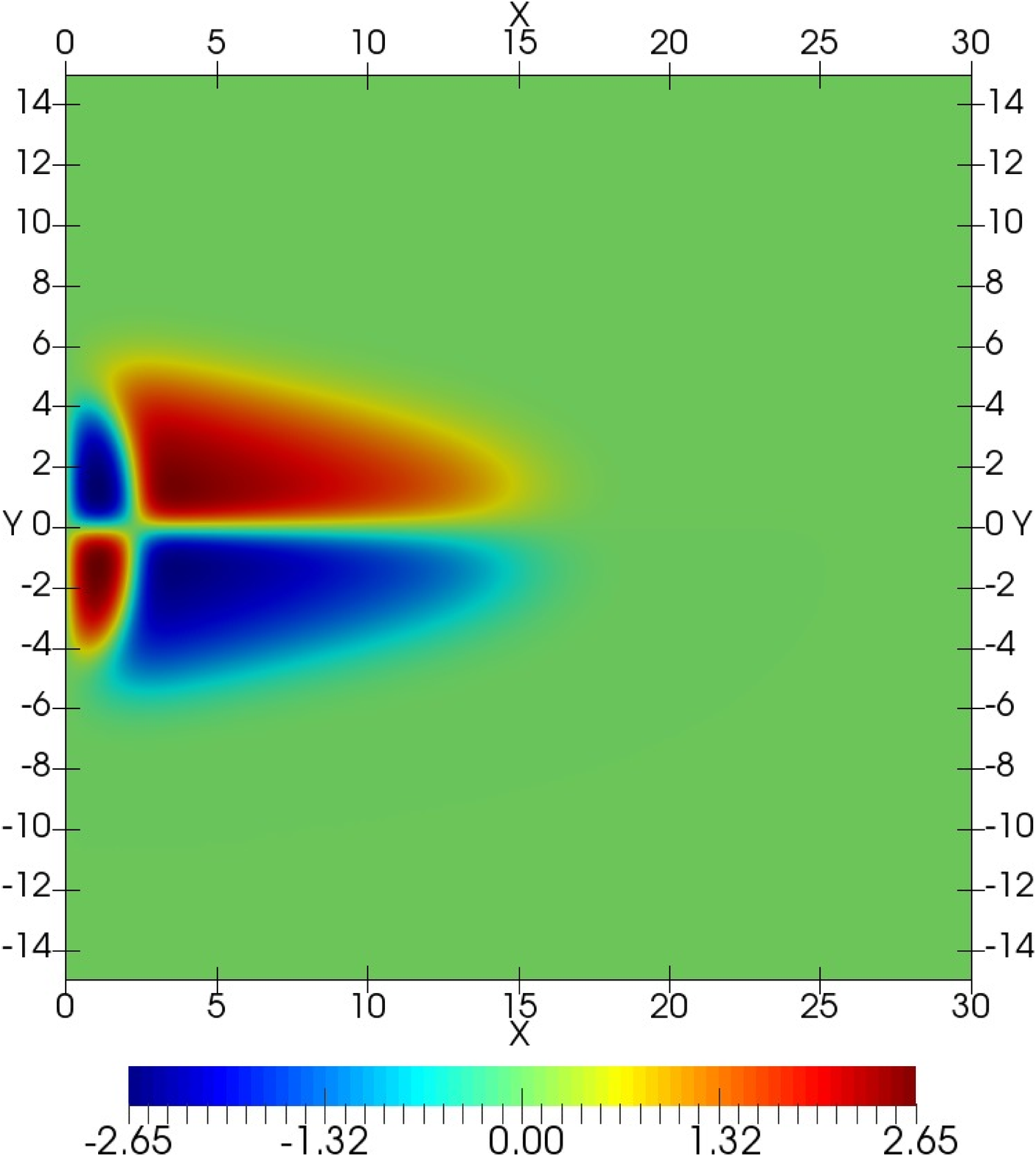}
\caption{$\mu=9.1$, $N\approx 318.97$, \\ \texttt{QN1\_x=1, QN1\_y=1}}
\end{subfigure}
\qquad
\begin{subfigure}{5cm} 
\includegraphics[width=5cm, keepaspectratio=true]{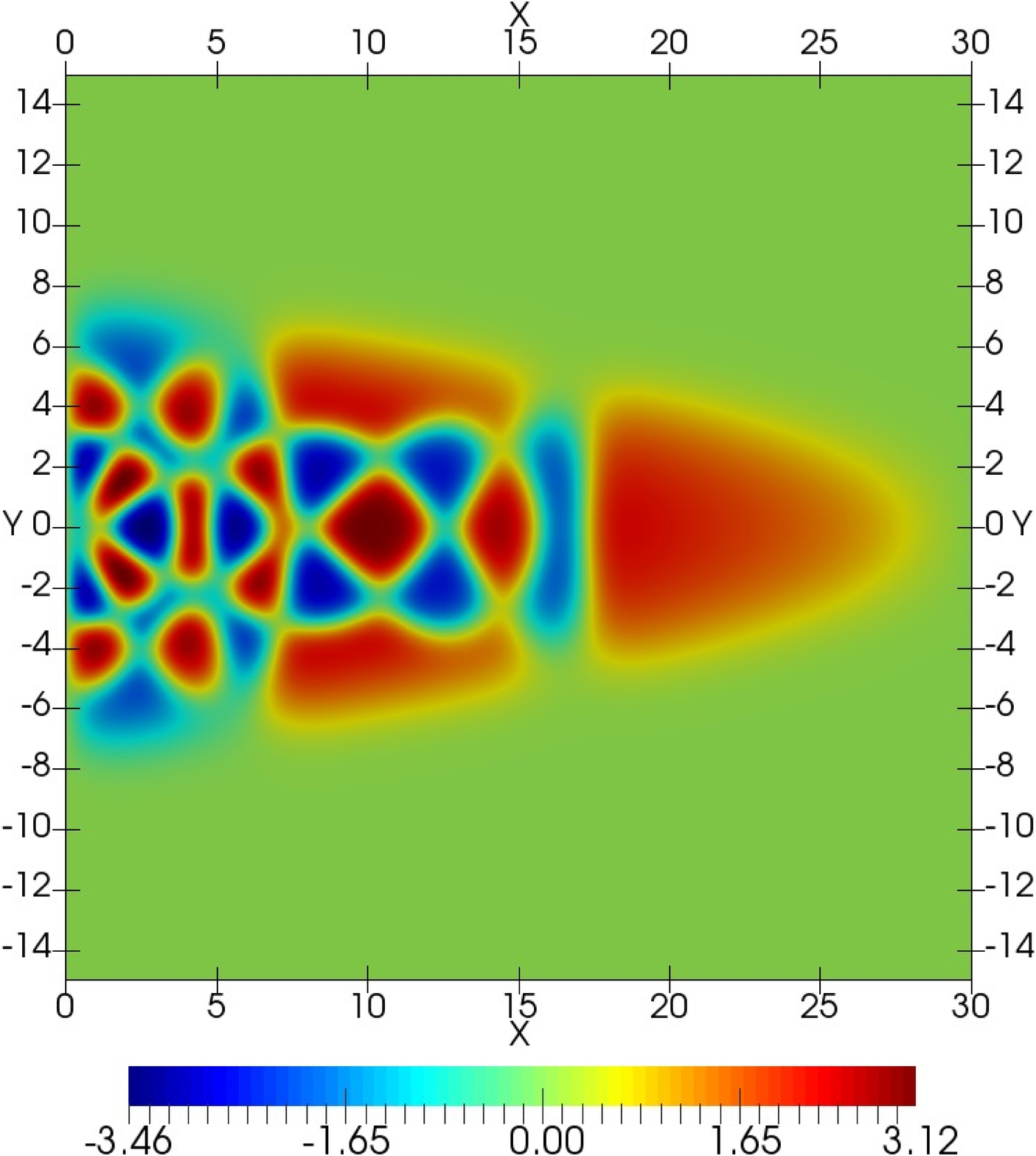}
\caption{$\mu=14.6$, $N\approx 647.13$, \\ \texttt{QN1\_x=4, QN1\_y=4}}
\end{subfigure}

\caption{\small Three different solution of the stationary GPE with $V(x,y) = x/2 + y^2/4$, $\gamma=1$. (a) This is the ground state depicted together with the default grid. (b) Results found by \texttt{breed} should look like this, if nothing is changed in the \texttt{params.prm} file. (c) An exemplary solution with a complicated structure.}\label{plot:gost}
\end{figure}

\begin{figure}[h]
\begin{subfigure}{5cm} 
\includegraphics[width=5cm, keepaspectratio=true]{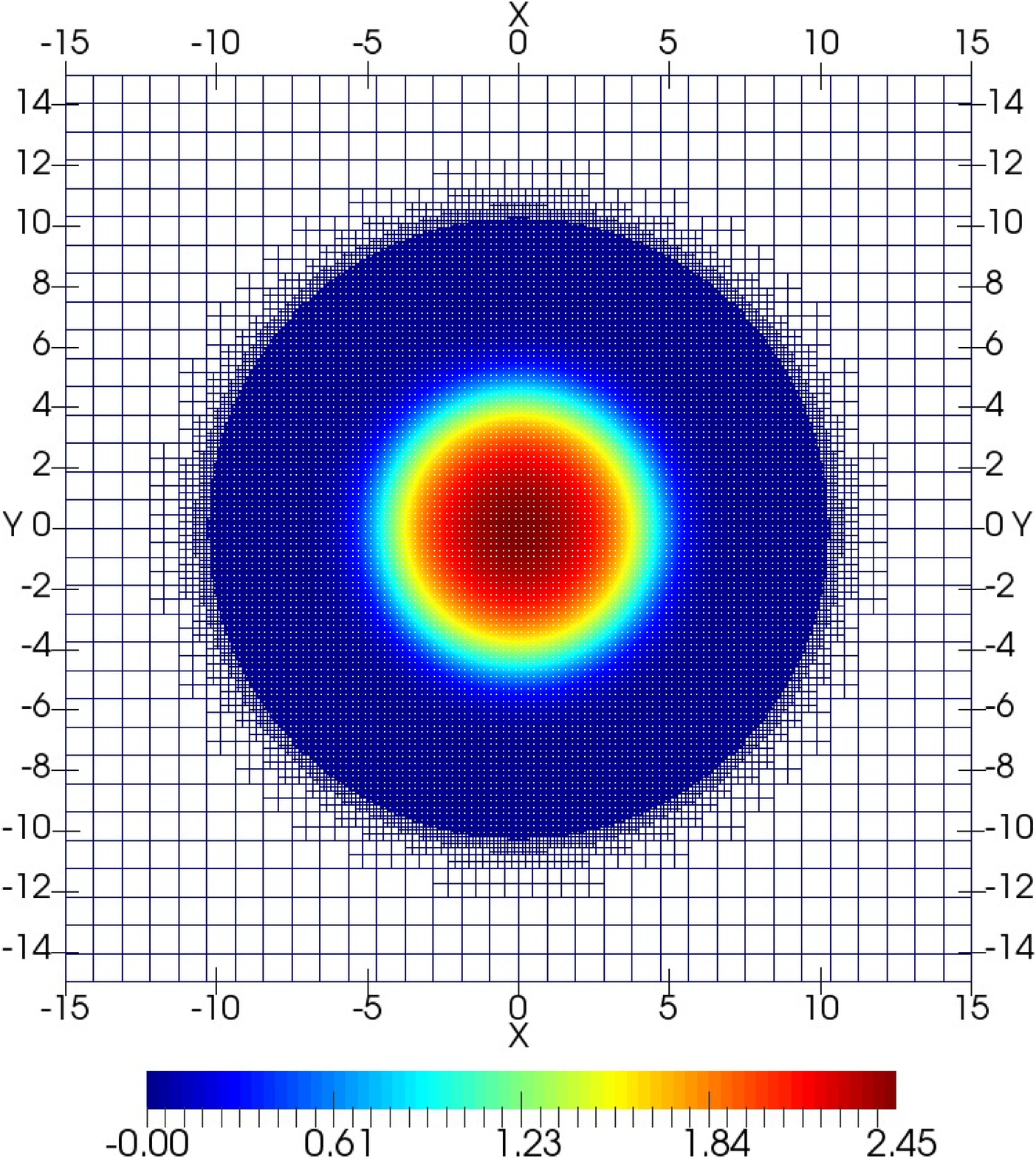}
\caption{$\mu=6.1$, $N\approx 225.25$, \\ \texttt{QN1\_x=0, QN1\_y=0}}
\end{subfigure}
\qquad
\begin{subfigure}{5cm} 
\includegraphics[width=5cm, keepaspectratio=true]{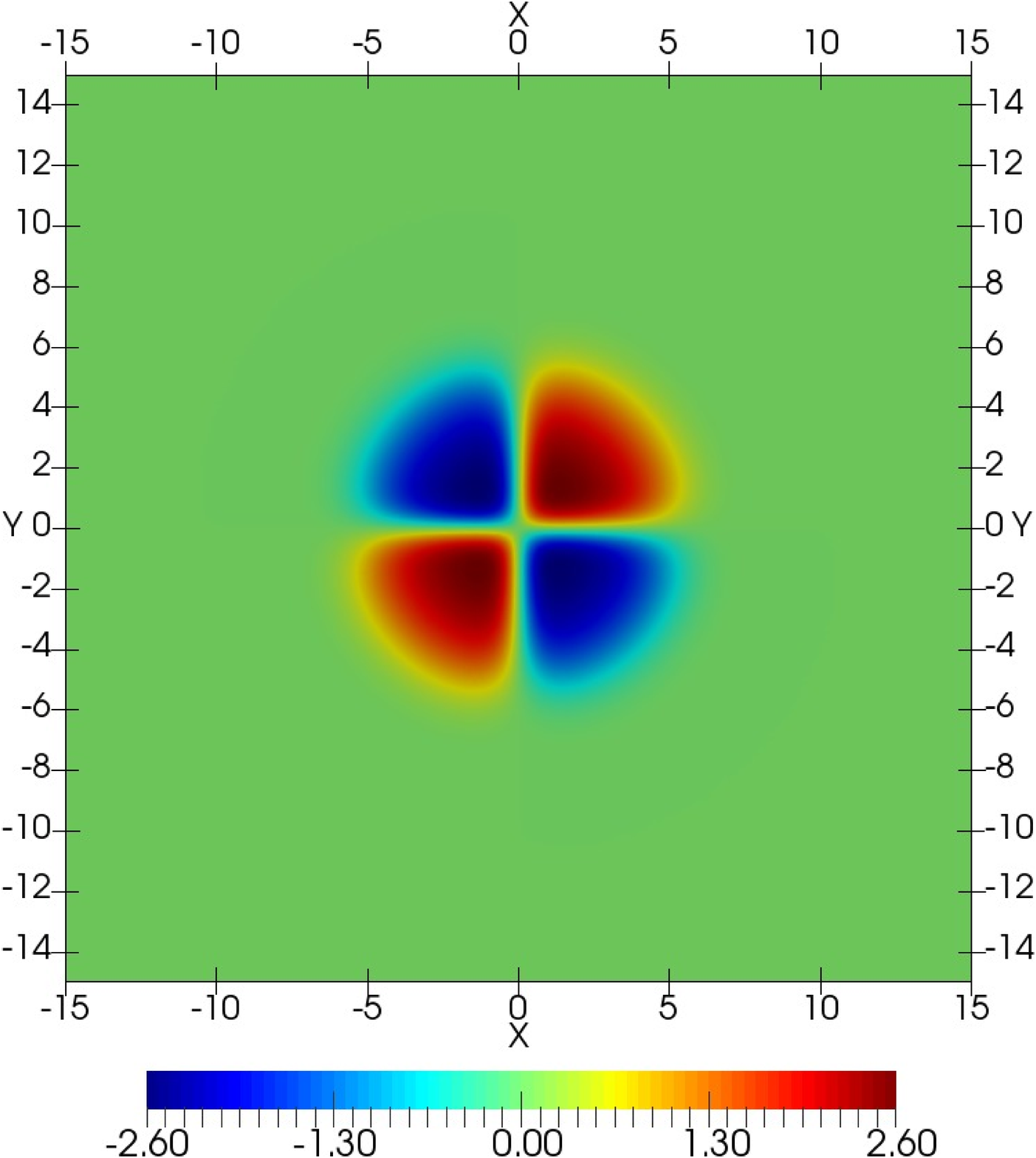}
\caption{$\mu=8.1$, $N\approx 265.76$, \\ \texttt{QN1\_x=1, QN1\_y=1}}
\end{subfigure}
\qquad
\begin{subfigure}{5cm} 
\includegraphics[width=5cm, keepaspectratio=true]{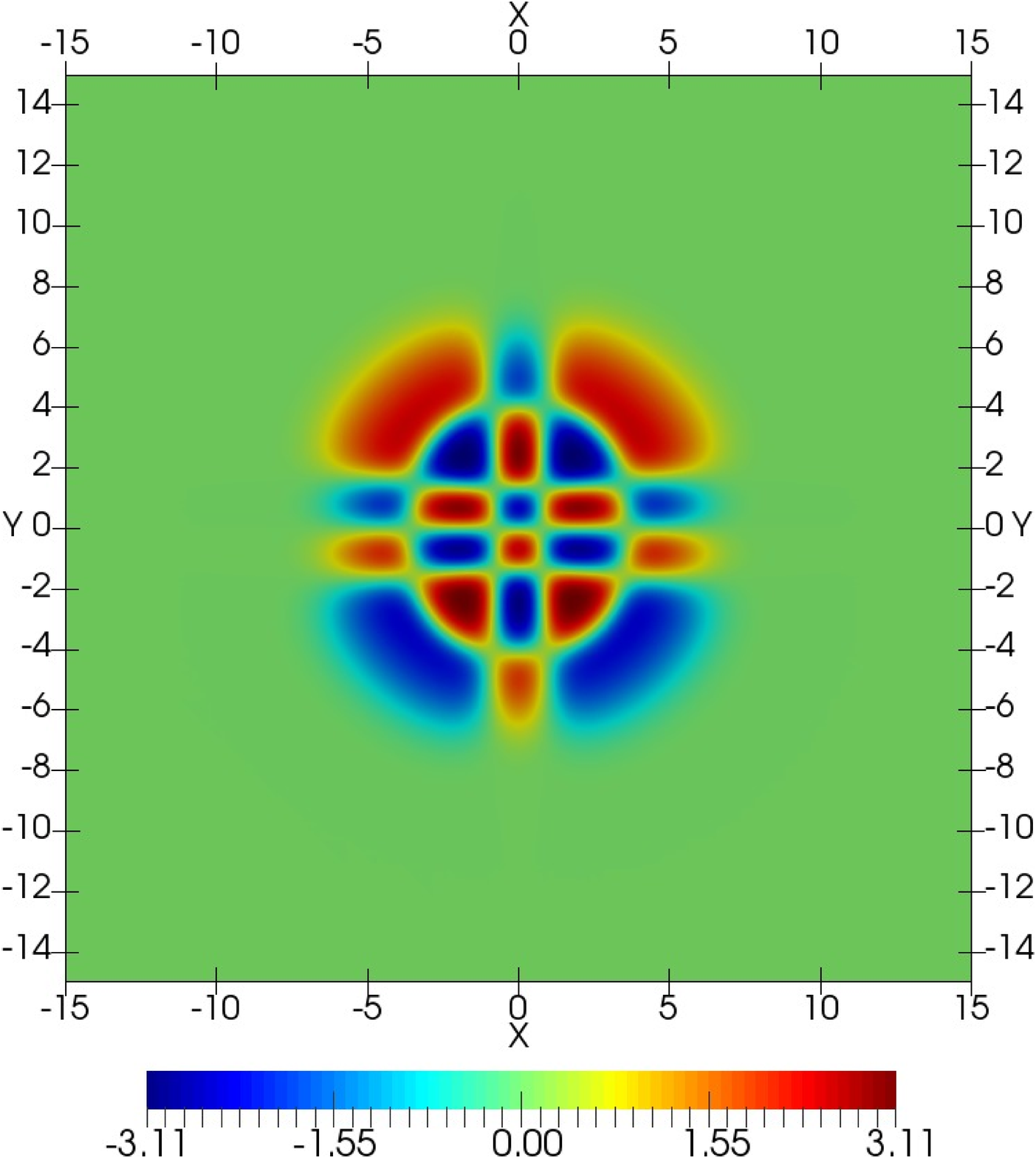}
\caption{$\mu=13.1$, $N\approx 378.14$, \\ \texttt{QN1\_x=4, QN1\_y=3}}
\end{subfigure}

\caption{\small Three different solution of the stationary GPE with $V(x,y) = x^2/4 + y^2/4$, $\gamma=1$. a) This is the ground state depicted together with the default grid. (b) In case of \texttt{BUILD\_HTRAP=OFF} a solution of a default run \texttt{breed} should look like this. (c) An exemplary solution with a complicated structure.}\label{plot:ho}
\end{figure}

\section{Example of a test-run}\label{TESTRUN}

\subsection{Stationary states}
A test-run may be performed within any directory by simply executing \texttt{mpirun -np [no of CPU cores] breed}. In case that the parameter file \texttt{params.prm} is missing, this will automatically generate one for a default scenario which is a gravito-optical surface trap in two dimensions (see equation (\ref{GOST})). Then, before starting the calculations, \texttt{breed} creates ten subfolders which are sorted in ascending order with respect to the chemical potential $\mu$. The value of each $\mu$ can be looked up in the file \texttt{log.csv} in each of the subfolders.  Note that the simulation then runs immediately and generates stationary states in Cartesian coordinates for each of the ten predetermined values of $\mu$. Here, the initial guess functions are chosen with quantum numbers $(n_x,n_y)=(1,1)$.  Finally, the results are written into each of the mentioned subfolders into the files \texttt{final.bin}, \texttt{final.bin.info} and \texttt{final-00001.vtu} (these are the default names). The latter serves for visualisation purposes and can be read, for example, by ParaView. As a visual example for this test-run, please see figures \ref{plot:gost} and \ref{plot:ho}.

\subsection{Dynamics}
The previously generated stationary states can serve as initial states for the simulation of the time-evolution.  For this, \texttt{mpirun -np [no of CPU cores] rtprop} can be initiated in any of the subfolders, where the files \texttt{final.bin}, \texttt{final.bin.info} and \texttt{params.prm} are present. In the default setup, the simulation runs until $t_{total}=1$ (dimensionless units) in steps of $\texttt{dt}=1e-2$ has been reached, where after each time step $\Delta t=0.1$ an output file is written. This can be adjusted in the parameter file \texttt{params.prm} by means of the parameter \texttt{NA} accounting for the frequency of data output. In the case of this test-run $t_{total}=1$ and $\texttt{NA}=10$ which translates to $t_{total}\cdot \texttt{NA}= 10$ output files in total. Another parameter which is adjusted in \texttt{params.prm} is \texttt{NK} specifies the number of intermediate steps between two file outputs. The results of the intermediate steps are printed in the terminal from which the simulation has been started. In the test-run the default value is $\texttt{NK}=10$, therefore a total number of $\texttt{NA} \cdot \texttt{NK}= 100$ steps is performed in the simulation, each with time step $\texttt{dt}=1e-2$. Thus, the total running time $t_{total}$ is determined by the product $t_{total} = \texttt{NA} \cdot \texttt{NK} \cdot \texttt{dt}$.

\section{Conclusions and Outlook}
We have presented the C++ package \texttt{atus-pro} for computation of solutions of the stationary as well as the time-dependent Gross-Pitaevskii equation. The code incorporates finite element methods in order to speed up calculations for BECs with complex spatial structures. These could be invoked by excited states solutions, e.g. high oscillation quantum numbers of the guess function, as well as by elaborated three-dimensional external potential setups. 

Here we would like to point out briefly the main differences between our code and those described in \cite{Muruganandam20091888} and \cite{Vudragovic20122021}. For the solution of the non-stationary case we use the fully implicit Crank-Nicolson method, leading to a system of non-linear equations. These are solved with a standard Newton method. In comparison, in the abovementioned works the semi-implicit Crank-Nicolson scheme is used and an operator splitting is carried out. This is the Alternating-Direction Implicit (ADI) method where the Laplace operator is split. Furthermore, this has been implemented for finite differences, whereas we use finite element methods.  
Finally, and this is the main difference, our stationary solver allows for calculation of excited states solutions.  

Excited topological modes have been also considered in \cite{yukalov_numsim_2015}, where a damping gradient method is used. We like to point out in the following, how that method differs from our constrained Newton algorithm.      
Finding the ground state in the non linear case with gradient based methods (Quasi-Newton) should work always, but finding saddle point solutions (topological modes, non-ground state solutions) becomes difficult for high non linearities and high excitations. 
Finding the first or second excited state is possible with a good Quasi-Newton method with a good initial guess. The mathematical reason is that the Morse index increases for complicated solutions. The Morse index is the number of negative eigenvalues of the second variational derivative of equation (\ref{gpe:functional}). This basically provides the information how many linearly independent descent directions are available.
In the numerical world one can only deal with approximated solutions, which lie near critical points. Therefore, gradient based methods tend to miss critical points if the Morse index is too high.
Sometimes they even simply miss it. The idea of the contrained Newton method goes back to Mountain pass or more general, to the Linking theorem (see the references in \cite{marojevic_energy_2013}). The search is restricted to a special manifold and the saddle points lie on it. We cannot give a guarantee that all saddle points can be found, however we think that more solutions can be found than using other methods. This does not take into account numerical continuation methods. 

Numerical continuation is able to find more solutions, but there it is necessary to detect bifurcation points by evaluating, roughly speaking, the change in the Morse index. This is very time consuming for large 
problems in more then one spatial dimension. So at the end a even more powerful solver would incorporate a combination of both methods, that is including a constrained Newton method. 
A possible future extension is to implement numerical continuation. The aim of numerical continuation is to follow solution branches
and detect possible branch points. Generally, a solution needs to be known beforehand. A Matlab tool box for numerical continuation which is based on this solution strategy is \texttt{pde2path} \cite{uecker_pde2path}. However, for our \texttt{breed} solvers it is not required to start with a known solution. In fact our code uses eigenfunctions of the corresponding linear problem just for convenience reasons.
One could think of using the \texttt{breed} solvers to find more complicated solutions without knowing the branch. That is,  by incorporating initial guess functions having a complex spatial structure. As suggested in \cite{kuehn_efficient_2014} this can be utilised as a starting point for numerical continuation. In contrast to other procedures, it is then not necessary to detect branching points in order to find new branches. 

To enhance the range of physical applications of our code, an extension to dual-species BEC setups is intended. This is especially important in the frame of modelling experiments which test the Universality of Free Fall (see, e.g. \cite{schlippert_dualspeciesUFF}). For this, an already existing two-species code will be adapted to the \texttt{deal.ii} library.   

Concerning technical extensions to the \texttt{atus-pro} code, it should be straightforward to extend for periodic boundary conditions. Furthermore, in order to be able to take into account a greater variety of domain geometries, our codes could be enhanced to incorporate the loading of grids, which can be generated by e.g.  \texttt{Gmsh}   or \texttt{Open CASCADE}. 

Finally, the implementation of adaptive grid refinement is envisaged.

\section{Acknowledgements}
We thank Dr. C. Pfeifer and K. Kanevche for careful reading, suggestions, testing of the installation instructions and validating the test-runs. Also we would like to thank Prof. H. Uecker from the University of Oldenburg for fruitful discussions.

We also would like to mention the deal.ii user group, which has been very helpful in solving problems accurately - especially for providing us with the mentioned patch for 32-bit systems. Moreover, we recommend the excellent online lectures of W. Bangerth, which can be found on \url {http://www.math.tamu.edu/~bangerth/videos.html}. 

Furthermore, we would like to emphasise that this work benefited from the inspiring environment and contact with the people of the research training group ``Models of Gravity'', which is supported by the German Research Foundation (DFG). 

This project is supported by the German Space Agency (DLR) with funds provided by the Federal Ministry of Economics and Technology (BMWi) due to an enactment of the German Bundestag under Grant No. DLR  50WM1342.

\begin{appendix}
 \section{Installation of required libraries}
Since the \texttt{atus-pro} package depends on many extra libraries which have to be compiled, too, and this process sometimes can get  cumbersome, we will give detailed compiling and installation instructions.  
\\

The packages needed for a successful installation of our program code can be divided into two groups: the first contains packages which are included in the most common Linux distributions, namely 
\paragraph{1st Group}
\begin{itemize}
\item CMake (min 2.8)
\item Doxygen (optional)
\item Python (2.7), Bison, Flex
\item C, C++ 11 compliant compiler, Fortran
\item zlib (including also the development package)
\end{itemize}
whereas in the second group software is listed, which - usually - has to be retrieved from external sources:
\paragraph{2nd Group}
\begin{itemize}
\item MPI Implementation
\item GSL - GNU Scientific Library (1.16)
\item LAPACK (3.5.0)
\item P4EST (1.1)
\item PETSC (3.6.1)
\item deal.II (8.3.0).
\end{itemize}

\subsection{General remarks on installation of dependencies}
As already mentioned, the packages of the 1st Group should be contained and be easily accessible in all common distributions of Linux. For those, no compiling is necessary, so we will only give installing instructions. On Ubuntu you can use the following command 
\begin{lstlisting}
sudo apt-get install g++ gfortran cmake cmake-curses-gui bison flex doxygen.
\end{lstlisting} 
On Redhat (OpenSUSE) like Linux distributions you may use 
\begin{lstlisting}
sudo yum (zypper) install gcc-c++ gcc-fortran doxygen cmake zlib-devel flex bison.
\end{lstlisting}

In the following, we focus on the installing instructions for the packages listed in the 2nd Group. We assume that you possess root privileges in order to be able to install all dependencies in \texttt{/opt}. But of course you are free to install everything into your local home folder by adding \texttt{\~} in front of \texttt{/opt}. 

For convenience, you should be able to copy and paste the commands from the boxes.
In addition, the file \texttt{INSTALL\_OPTIONS.txt} is included in the archive  \texttt{atus-pro\_v1.0.tar.gz}, where the install and compilation options for all the packages are also listed and can be copied to a terminal from which installation takes place.  

We suggest to extract all the downloaded packages listed in \ref{MPI}, \ref{GSL}, \ref{LAPACK}, \ref{P4EST}, \ref{PETSC} and \ref{DEALII} to a temporary folder. In the following we assume that the folder name is \texttt{temp} and located in the top level of your home directory, i.e. the complete path would be then \texttt{\$HOME/temp}.  

\subsection{Installation of MPI}\label{MPI}

You can skip this step if you have an MPI implementation already installed, which has also Fortran enabled. Here we present the installation of Open MPI, but other implementations of the MPI standard should work as well. 

It is important to check the \texttt{lib} folder naming of each library after each installation. Depending on your Linux distribution this may differ. It can be labelled as \texttt{lib} or \texttt{lib64} on 64-bit operating systems, whereas on 32-bit operating systems it should be always named as \texttt{lib}. You may check this via the command \texttt{uname -a} beforehand. Note that a wrong \texttt{LD\_LIBRARY\_PATH} will cause a failure of the compilation process.

\begin{lstlisting} 
(0) cd $HOME # start from top level of your home folder
(1) mkdir temp # create temporary folder 
(2) cd temp
(3) wget http://www.open-mpi.org/software/ompi/v1.8/downloads/openmpi-1.8.5.tar.gz
(4) tar xfvz openmpi-1.8.5.tar.gz
(5) cd openmpi-1.8.5
(6) ./configure --prefix=/opt/openmpi-1.8.5 --enable-mpi-fortran
(7) make -j 
(8) sudo make install
(9) export PATH=$PATH:/opt/openmpi-1.8.5/bin
(10) export LD_LIBRARY_PATH=$LD_LIBRARY_PATH:/opt/openmpi-1.8.5/lib64
 # alternativly: export LD_LIBRARY_PATH=$LD_LIBRARY_PATH:/opt/openmpi-1.8.5/lib
(11) check the path with: which mpicc
(12) check the lib folder naming
\end{lstlisting} 

\subsection{Installation of GSL}\label{GSL}
The GNU Scientific Library \cite{gough_gnu_2009}, or short GSL, is required for the ansatz functions and for the search of the reference point in \texttt{breed} and \texttt{breed\_cs}. \\

\begin{lstlisting} 
(0) cd $HOME/temp
(1) wget ftp://ftp.gnu.org/gnu/gsl/gsl-1.16.tar.gz
(2) tar xfvz gsl-1.16.tar.gz 
(3) cd gsl-1.16
(4) ./configure --prefix=/opt/gsl-1.16
(5) make CFLAGS="-march=native -O3" # here we overload the default CFLAGS options
(6) sudo make install 
\end{lstlisting} 

\subsection{Installation of LAPACK}\label{LAPACK}
Lapack is needed by \texttt{deal.ii} for the single core direct solver called by the programs for the one-dimensional simulation \texttt{breed\_1} and \texttt{rtprop\_1}. Other available LAPACK versions may also work, but they have not been tested so far.

\begin{lstlisting} 
(0) cd $HOME/temp
(1) wget http://www.netlib.org/lapack/lapack-3.5.0.tgz
(2) tar xfvz lapack-3.5.0.tgz
(3) cd lapack-3.5.0
(4) mkdir build
(5) cd build
(6) cmake -DCMAKE_BUILD_TYPE=Release -DCMAKE_Fortran_FLAGS_RELEASE="-march=native -fpic -O3" \
    -DCMAKE_INSTALL_PREFIX=/opt/lapack-3.5.0  ..
(7) make -j
(8) sudo make install
\end{lstlisting} 

\subsection{Installation of p4est}\label{P4EST}
The p4est library requires MPI. The p4est library \cite{BangerthBursteddeHeisterEtAl11} is responsible for the management of the parallel triangulation.
It is important to mention that from now on the order of installation of the last three packages (including this one) is important and has to be performed in the order of the numbering of the subsections, i.e.  
install \ref{P4EST}, \ref{PETSC} and finally \ref{DEALII}. 

\begin{lstlisting} 
(0) cd $HOME/temp
(1) wget http://p4est.github.io/release/p4est-1.1.tar.gz
(2) tar xfvz p4est-1.1.tar.gz
(3) cd p4est-1.1
(4) ./configure --prefix=/opt/p4est-1.1 --enable-mpi --enable-shared --disable-vtk-binary --without-blas
(5) make CFLAGS="-O3 -march=native"
(6) sudo make install
\end{lstlisting}

\subsection{Installation of PETSc}\label{PETSC}

The PETSc \cite{petsc-web-page,petsc-user-ref,petsc-efficient} library requires MPI and a working internet connection. The PETSc library provides the deal.II library with routines which are required to solve systems of linear equations in parallel. We prefer the parallel direct solver MUMPS (\url{http://mumps.enseeiht.fr}). 

\begin{lstlisting} 
(0) cd $HOME/temp
(1) wget http://ftp.mcs.anl.gov/pub/petsc/release-snapshots/petsc-3.6.1.tar.gz
(2) tar xfvz petsc-3.6.1.tar.gz
(3) cd petsc-3.6.1
(4) export PETSC_ARCH=x86_64 # you can choose any arbitrary string you like
(5) ./configure --prefix=/opt/petsc-3.6.1 --with-shared-libraries --with-x=0 --with-debugging=0 \
    --with-mpi=1 --download-hypre=yes --download-fblaslapack=1 --download-scalapack --download-mumps \
    --download-ptscotch 
(6) make PETSC_DIR=$HOME/temp/petsc-3.6.1 PETSC_ARCH=x86_64 all
(7) sudo make PETSC_DIR=$HOME/temp/petsc-3.6.1 PETSC_ARCH=x86_64 install
\end{lstlisting} 

\subsection{Installation of deal.II}\label{DEALII}
The deal.II library requires MPI and the preceding libraries. In order to speed up the compilation process you can optionally invoke \texttt{ccmake .} from within the folder \texttt{\$HOME/temp/dealii-8.3.0/build} (see the box below) and change the entry in CMAKE\_BUILD\_TYPE to \texttt{Release}.

Moreover, if you want to benefit from an extra speed-up you can utilise  more (or all) available CPU cores through executing \texttt{make -j[number of CPU cores]} in step (7). As a precaution, we do not recommend using just \texttt{make -j} (i.e. without explicitly determining the number of CPU cores) since the build process may slow down your computer extremely !

\begin{lstlisting} 
(0) cd $HOME/temp
(1) wget https://github.com/dealii/dealii/releases/download/v8.3.0/dealii-8.3.0.tar.gz
(2) tar xfvz dealii-8.3.0.tar.gz
(3) cd dealii-8.3.0
(4) mkdir build
(5) cd build
(6) cmake -DDEAL_II_WITH_UMFPACK=ON -DDEAL_II_WITH_LAPACK=ON -DLAPACK_DIR=/opt/lapack-3.5.0 \
    -DPETSC_ARCH=x86_64 -DPETSC_DIR=/opt/petsc-3.6.1 -DP4EST_DIR=/opt/p4est-1.1 \ 
    -DDEAL_II_WITH_THREADS=OFF -DDEAL_II_WITH_MPI=ON -DDEAL_II_WITH_HDF5=OFF \ 
    -DCMAKE_INSTALL_PREFIX=/opt/deal.II-8.3.0 .. 
# after cmake has finished, it prints a summary of your configuration. The configuration should look
# like the output in the box below. 
(7) make # alternatively try make -j[no of cpu cores] 
(8) sudo make install
\end{lstlisting}
After step (7), upon successful compilation, the terminal output should look like this for the user ``johndoe'':
\begin{lstlisting} 
###
#
#  deal.II configuration:
#        CMAKE_BUILD_TYPE:       DebugRelease
#        BUILD_SHARED_LIBS:      ON
#        CMAKE_INSTALL_PREFIX:   /opt/deal.II-8.3.0
#        CMAKE_SOURCE_DIR:       /home/johndoe/temp/dealii-8.3.0
#                                (version 8.3.0)
#        CMAKE_BINARY_DIR:       /home/johndoe/temp/dealii-8.3.0/build
#        CMAKE_CXX_COMPILER:     GNU 4.8.3 on platform Linux i686
#                                /usr/bin/c++
#
#  Configured Features (DEAL_II_ALLOW_BUNDLED = ON, DEAL_II_ALLOW_AUTODETECTION = ON):
#      ( DEAL_II_WITH_64BIT_INDICES = OFF )
#      ( DEAL_II_WITH_ARPACK = OFF )
#        DEAL_II_WITH_BOOST set up with external dependencies
#      ( DEAL_II_WITH_BZIP2 = OFF )
#        DEAL_II_WITH_CXX11 = ON
#      ( DEAL_II_WITH_CXX14 = OFF )
#      ( DEAL_II_WITH_HDF5 = OFF )
#        DEAL_II_WITH_LAPACK set up with external dependencies
#      ( DEAL_II_WITH_METIS = OFF )
#        DEAL_II_WITH_MPI set up with external dependencies
#        DEAL_II_WITH_MUPARSER set up with bundled packages
#      ( DEAL_II_WITH_NETCDF = OFF )
#      ( DEAL_II_WITH_OPENCASCADE = OFF )
#        DEAL_II_WITH_P4EST set up with external dependencies
#        DEAL_II_WITH_PETSC set up with external dependencies
#      ( DEAL_II_WITH_SLEPC = OFF )
#      ( DEAL_II_WITH_THREADS = OFF )
#      ( DEAL_II_WITH_TRILINOS = OFF )
#        DEAL_II_WITH_UMFPACK set up with bundled packages
#        DEAL_II_WITH_ZLIB set up with external dependencies
#
#  Component configuration:
#      ( DEAL_II_COMPONENT_DOCUMENTATION = OFF )
#        DEAL_II_COMPONENT_EXAMPLES
#        DEAL_II_COMPONENT_MESH_CONVERTER
#      ( DEAL_II_COMPONENT_PACKAGE = OFF )
#      ( DEAL_II_COMPONENT_PARAMETER_GUI = OFF )
#
#  Detailed information (compiler flags, feature configuration) can be found in detailed.log
#
#  Run  $ make info  to print a help message with a list of top level targets
#
###
-- Configuring done
-- Generating done
-- Build files have been written to: /home/johndoe/temp/dealii-8.3.0/build
johndoe@linux:~/temp/dealii-8.3.0/build>  
\end{lstlisting}  

\section{Building the \texttt{atus-pro} package}

Before we can build the \texttt{atus-pro} package from the source it is necessary to setup the environmental variables. The \texttt{PATH} variable should point to the folder where the MPI binaries are located. In order to be capable to run our programs, in addition, it must include \texttt{\$HOME/bin}. The \texttt{LD\_LIBRARY\_PATH} variable should contain all paths pointing to all library locations. Our CMake configuration uses this path to search for the library files. 

We suggest to create a file (e.g. \texttt{.my\_bashrc}) located in the home folder \texttt{\$HOME} with the contents given in the box below. This file can then be used to set up the paths by invoking the \texttt{source} command, i.e. \texttt{source \$HOME/.my\_bashrc}.  In general, we recommend the use of environment modules (\url{http://modules.sourceforge.net/}) which is common on distributed memory machines.

\begin{lstlisting}[caption=.my\_bashrc]
export PATH=$PATH:$HOME/bin:/opt/openmpi-1.8.5/bin
export LD_LIBRARY_PATH=$LD_LIBRARY_PATH:/opt/openmpi-1.8.5/lib64:/opt/gsl-1.16/lib64:/opt/p4est-1.1/lib64: \
                       /opt/petsc-3.6.1/lib:/opt/deal.II-8.3.0/lib
# alternatively
export LD_LIBRARY_PATH=$LD_LIBRARY_PATH:/opt/openmpi-1.8.5/lib:/opt/gsl-1.16/lib:/opt/p4est-1.1/lib: \
                       /opt/petsc-3.6.1/lib:/opt/deal.II-8.3.0/lib
                       
\end{lstlisting}
At this point we remind you to check the \texttt{lib} folder naming above in \texttt{LD\_LIBRARY\_PATH}. On 64-bit operating system it may be \texttt{lib} or \texttt{lib64}, whereas on 32-bit operating systems it should be always named \texttt{lib} ! For this, it is recommended to check the real installation location of the libraries listed in \texttt{LD\_LIBRARY\_PATH}. Note that by using the command \texttt{uname -a} you can verify if you are using a 32- or 64-bit operating system.    

Before starting to compile \texttt{atus-pro}, we assume that you have already copied the archive \texttt{atus-pro\_v1.0.tar.gz} into your \texttt{\$HOME/temp} folder !
\begin{lstlisting}
(0) source $HOME/.my_bashrc # establish the paths of the previously compiled libraries
(1) cd $HOME/temp
(2) tar xfvz atus-pro_v1.0.tar.gz
(3) cd atus-pro_v1.0
(4) mkdir build
(5) cd build 
(6) cmake ..
(7) make -j # after the build process you can find the binaries in $HOME/bin
(8) make doc # optionally: build the doxygen documentation  
\end{lstlisting}

Within the build folder you can view all CMake options and preferences through invoking the \texttt{ccmake .} command. If CMake has \texttt{doxygen} detected, then you can build the HTML documentation via the command \texttt{make doc}. The documentation is then accessible through the index.html located in the HTML folder. This documentation contains a more detailed description of the algorithms.

\subsection{Important CMake options}\label{CMake_options}
The CMake options can be accessed by executing \texttt{ccmake .} in the folder \texttt{\$HOME/temp/atus-pro\_v1.0/build}.
Editing the following listed options is necessary in case you decide to change the default settings (i.e. switch from 2D to 3D problems). In order for the changes to take effect, this must be done prior to compiling the \texttt{atus-pro} package.

If you build the package for the first time without changing these options, then the initial setup of the external potential is a two-dimensional gravito-optical surface trap in accordance with equations (\ref{GOST}) and (\ref{GOST_cs}). One can switch to a purely harmonic trap by setting \texttt{BUILD\_HTRAP ON}. The values of each trapping frequency $(\omega_x,\omega_y,\omega_z)$ are then fixed in the parameter file \texttt{params.prm}. Of course, if three-dimensional simulations are to be made, \texttt{BUILD\_3D} must be set to \texttt{ON} before compiling \texttt{atus-pro}.
\\
\begin{tabularx}{18cm}{|X|X|X|}
\hline 
CMake option & default& \\
 & value& \\
\hline 
\texttt{BUILD\_3D} & \texttt{OFF} & {If this option is set to ON, then our C++ templates are instantiated for three spatial dimensions.} \\
\hline 
\texttt{BUILD\_DOCUMENTATION} & \texttt{AUTO} & {If Latex and \texttt{doxygen} are available on your computer, then the build target for the documentation is automatically enabled .} \\
\hline 
\texttt{BUILD\_HTRAP} & \texttt{OFF} & {If this option is set to ON, then a harmonic trap is switched on.}  \\ 
\hline 
\texttt{BUILD\_NEHARI} & \texttt{ON} & {Influences the initial point in the function space in breed and breed\_cs. For a detailed explanation look into the \texttt{doxygen} documentation.} \\ 
\hline 
\texttt{BUILD\_VARIANT2} & \texttt{ON} & {Influences the search of the reference point in the function space in breed and breed\_cs. For a detailed explanation look into \texttt{doxygen} HTML documentation.} \\ 
\hline 
\end{tabularx} 

\subsection{Overview of parameters of the parameter file \texttt{params.prm}}

\begin{tabular}{|l|l|l|}
\hline
\texttt{NA} & Frequency of data output  \\
\hline
\texttt{NK} & Number of intermediate steps  \\
\hline
\texttt{Ndmu} & Number of $\Delta\mu$ steps  \\
\hline
\texttt{dmu} & $\Delta\mu$  \\
\hline
\texttt{df} & Damping factor for the Newton method  \\
\hline
\texttt{dt} & Time step $\Delta t$  \\
\hline
\texttt{epsilon} & Termination threshold for stationary solvers.   \\
\hline
\texttt{filename} & File name of the wave function for the initial time step  \\
\hline
\texttt{guess\_fct} & Set $\phi_0^0$ manually (see \texttt{doxygen} documentation) \\
\hline
\texttt{ti} & Initial value for point in function space. If \texttt{BUILD\_NEHARI} = \texttt{OFF}, $t_0^{s,k=0} \leftarrow \texttt{ti}$ and $t_1^{s,k=0} \leftarrow \texttt{ti}$  \\
\hline
\texttt{global\_refinements} & Level of global mesh refinements for the inital mesh  \\
\hline
\texttt{xMax} & Max. value of first coordinate of simulation box \\
\hline
\texttt{xMin} & Min. value of first coordinate of simulation box  \\
\hline
\texttt{yMax} & Max. value of second coordinate of simulation box  \\
\hline
\texttt{yMin} & Min. value of second coordinate of simulation box  \\
\hline
\texttt{zMax} & Max. value of third coordinate of simulation box  \\
\hline
\texttt{zMin} & Min. value of third coordinate of simulation box  \\
\hline
\texttt{QN1\_x} & Quantum number corresponding to the first coordinate  \\
\hline
\texttt{QN1\_y} & Quantum number corresponding to the second coordinate  \\
\hline
\texttt{QN1\_z} & Quantum number corresponding to the third coordinate  \\
\hline
\texttt{gf} & Gravitational acceleration  \\
\hline
\texttt{gs} & Self interaction parameter $\gamma$  \\
\hline
\texttt{omega\_x} & $\omega_x$  \\
\hline
\texttt{omega\_y} & $\omega_y$ or $\omega_r$  \\
\hline
\texttt{omega\_z} & $\omega_z$  \\
\hline
\texttt{t} & Starting time for the real time propagation  \\
\hline
\end{tabular}

\end{appendix}
\newpage

\bibliographystyle{elsarticle-num}
\bibliography{paper}

\end{document}